# Поиск красных карликов среди рентгеновских объектов глубокого обзора экваториальной области неба по данным eROSITA

*Алексей А. Шляпников*

**Абстракт**

В работе представлено отождествление звёзд нижней части Главной последовательности среди объектов глубокого обзора экваториальной области неба, выполненного телескопом eROSITA на орбитальной обсерватории SRG. Из 27910 рентгеновских объектов 110, возможно, являются кандидатами на идентификацию с красными карликами. 12 звёзд ранее были классифицированы, как рентгеновские источники. В области идентификации двух звёзд попадают галактики. Несколько отождествлений содержат близко расположенные объекты. Список идентифицированных объектов, комментарии к нему и поисковые карты даны в приложениях.

**Введение**

Рентгеновский телескоп eROSITA (extended ROentgen Survey with an Imaging Telescope Array – телескоп для широкоугольного обзора в рентгеновском диапазоне) [1] на борту обсерватории Спектр-Рентген-Гамма (Spectrum-Roentgen-Gamma - SRG) [2] обладает значительным полем (~ 1 градус в диаметре) для регистрации неба в диапазоне энергий от 0,2 до 8,0 кэВ. Телескоп также может выполнять наблюдения больших участков неба. Максимальный поддерживаемый размер прямоугольников с равномерным сканированием области составляет 12,5 x 12,5 градуса.

Обсерватория SRG была успешно запущена в июле 2019 года. До начала обзора всего неба, для проверки всех различных возможностей приборов, в начале миссии был проведён ряд специальных наблюдений.

Исследование экваториальной области неба (eFEDS - eROSITA Final Equatorial Depth Survey) стало самым длительным по времени наблюдением в период проверки возможностей приборов телескопа eROSITA. В целом было затрачено около 100 часов. Поле eFEDS имеет площадь примерно 140 град$^2$ и состоит из четырех отдельных прямоугольных областей сканирования по 35 град$^2$. Выбор площадки был обусловлен наличием значительного числа многоволновых наблюдений данной области неба [3].

**Поиск красных карликов в каталоге eFEDS**

По результатам глубокого обзора eFEDS было опубликовано два каталога рентгеновских источников: однозначно обнаруженные источники в диапазоне 0,2–2,3 кэВ и жесткие (2,3–5 кэВ) источники, наблюдавшиеся в многодиапазонном режиме [3]. В данной работе для идентификации объектов из нового Каталога звёзд с активностью солнечного типа CSAST [4] были выбраны 27910 рентгеновских источников с высокой степенью вероятности обнаружения, которые составляют основной каталог eFEDS [5].

Распределение на небесной сфере в проекции AITOFF звёзд из каталога CSAST и область объектов eFEDS продемонстрирована на рисунке 1. Каталог CSAST (версия 2.0 от 8 августа 2021 года) содержит 314618 объектов. Учитывая специфику CSAST, в него вошли звёзды из каталога GAIA DR2 [6], удовлетворяющие следующим требованиям:

$$T_\star < 7000° \text{ K} \qquad (1)$$
$$L_\star < 1.1 \text{ L}_\odot \qquad (2)$$
$$L_\star \geq 6.136^{-6} \times T_\star - 0.022 \qquad (3)$$

То есть, эффективная температура звёзд < 7000° K, светимость < 1.1 светимости Солнца. Неравенство 3 отсекает из каталога GAIA белые карлики, расположенные в нижней левой части диаграммы Герцшпрунга-Рассела.

Дополнительно применялся критерий отбора по радиусу. Он должен был не превышать 1.1 радиуса Солнца.

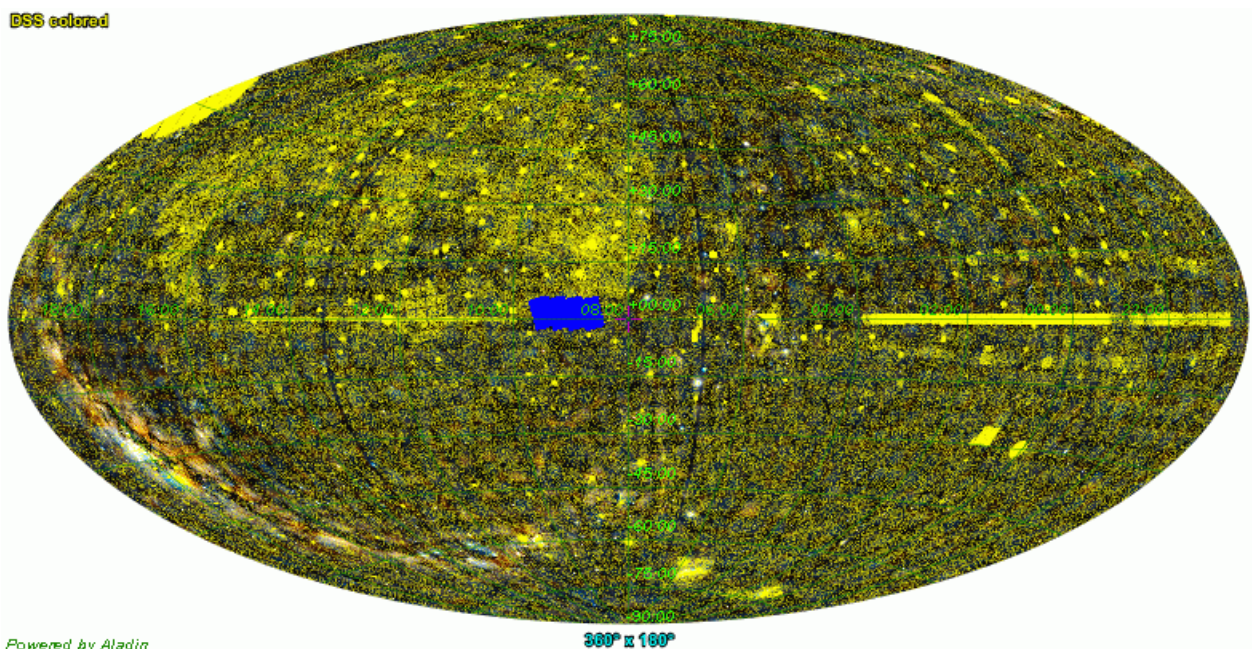

Рис. 1. Область eFEDS на фоне звёзд из каталога CSAST обозначена голубым цветом.

Более детальное распределение идентифицируемых объектов представлено на рисунке 2. На нём берюзовым цветом указаны объекты из каталога eFEDS, а желтыми маркерами обозначены звёзды из CSAST. Рисунок хорошо иллюстрирует четыре отдельные прямоугольные области сканирования eROSITA.

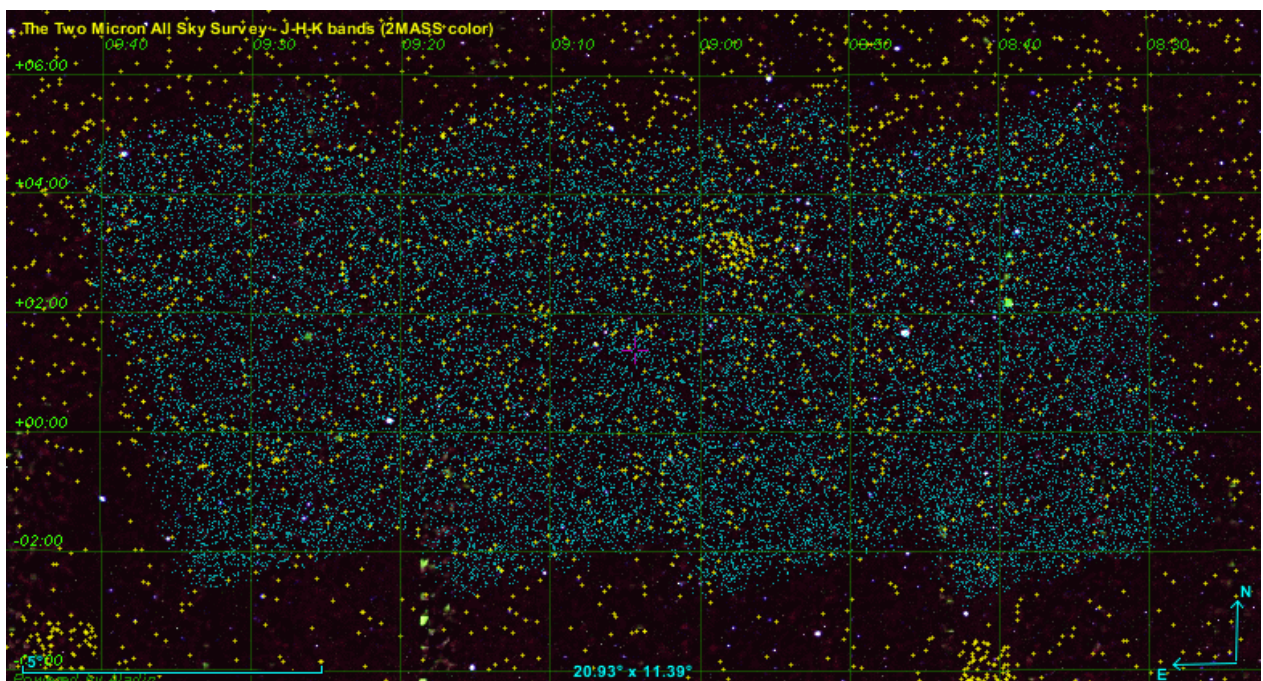

Рис. 2. Области eFEDS и звёзды из каталога CSAST (пояснения в тексте).

Для идентификации объектов из каталога eFEDS использовались откорректированные значения прямого восхождения и склонения объектов, а также комбинированная ошибка в определении координат. Рисунок 3 иллюстрирует гистограмму распределения числа ошибок в определении координат от их радиуса в угловых секундах для всех объектов eFEDS.

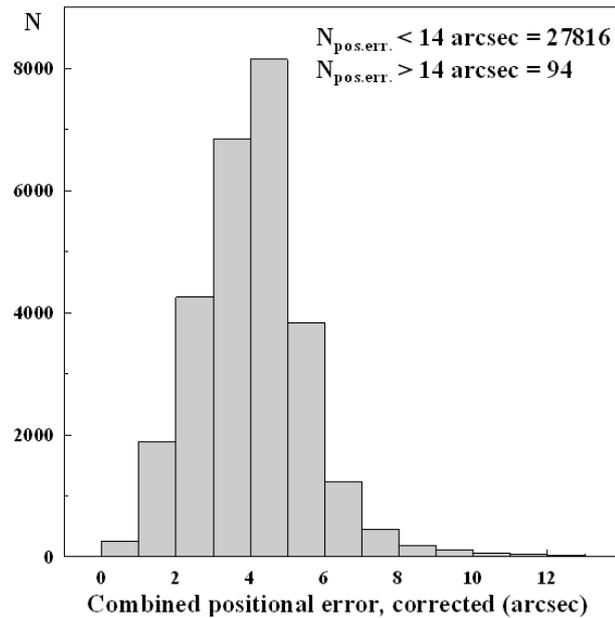

Рис. 3. Распределение числа ошибок в определении координат от их радиуса.

Учитывая, что средний радиус ошибок в определении координат eFEDS составляет 4.8 угловые секунды, поиск кандидатов на идентификацию производился в утроенном радиусе. На рисунках 4 и 5 показаны соответственно распределения: числа ошибок в определении координат для объектов – возможных кандидатов на отождествление с красными карликами и гистограмма числа объектов в зависимости от их расстояния до положения рентгеновского источника.

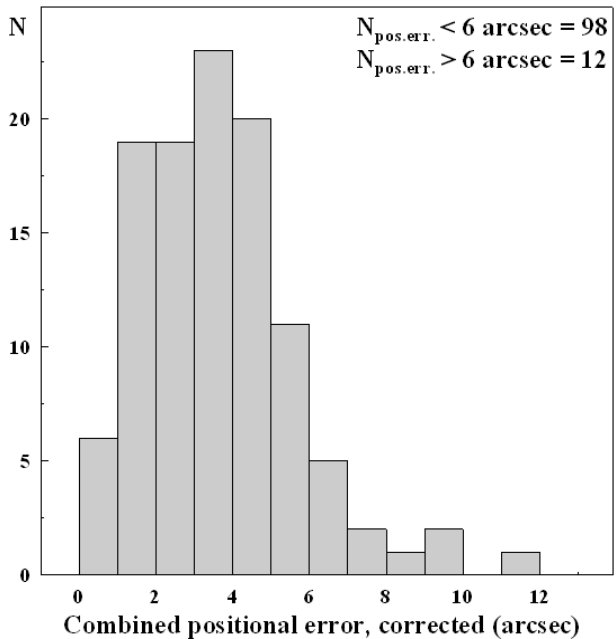

Рис. 4.

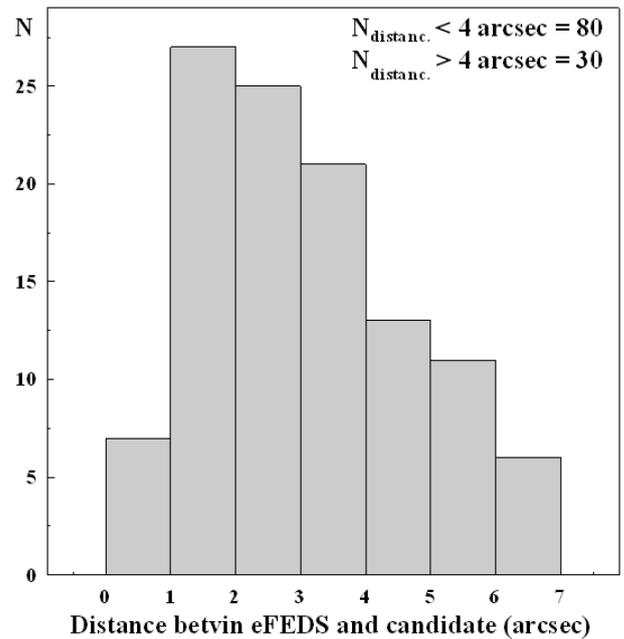

Рис. 5.

Рисунок 6 демонстрирует положение идентифицированных объектов на фоне распределения на небе рентгеновских источников из каталога eFEDS.

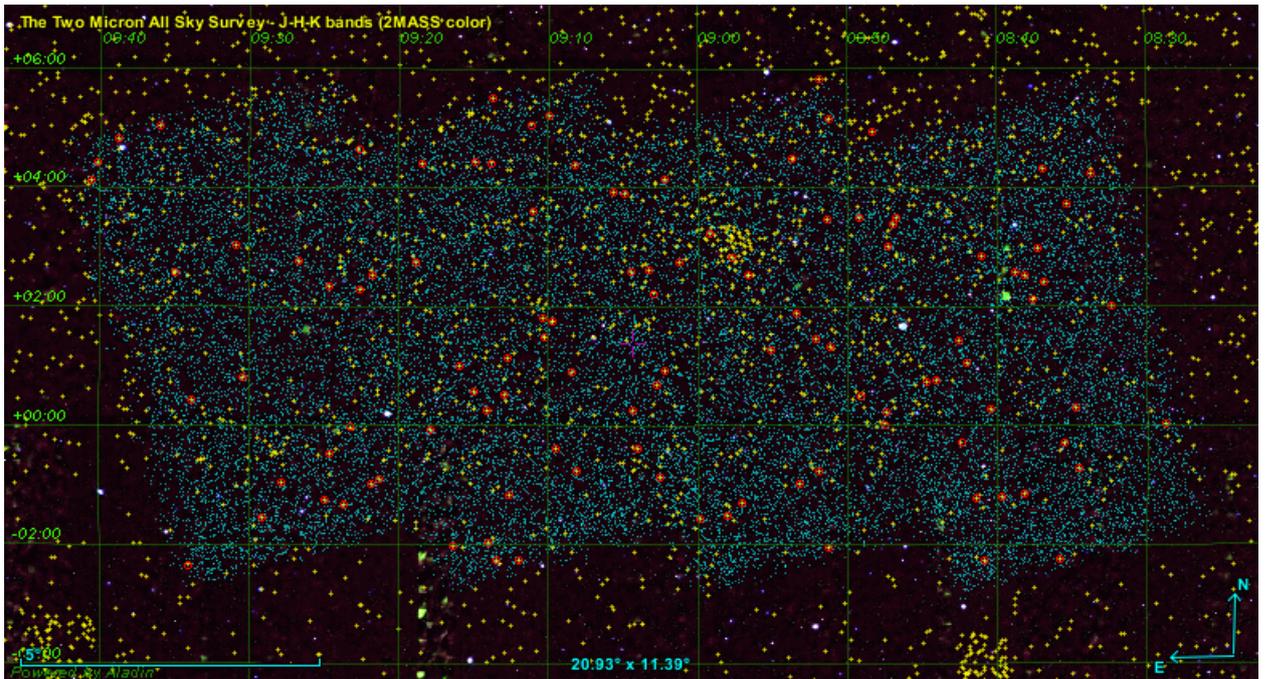

Рис. 6. Отождествлённые звёзды из каталога CSAST в областях eFEDS.

**Анализ идентифицированных объектов**

В таблице (Приложение 1) из 110 звёзд указаны две области, в которые попадают известные галактики: галактика Сейферта 1-го типа - SDSS J084818.23+045643.1 со звёздной величиной в полосе $G_{GAIA} = 20^m.306$ (рис. 7) и галактика SDSS J090807.97-004613.9 с редуцированной по данным GAIA звёздной величиной в полосе $V = 19^m.039$. Эта галактика находится на расстоянии 5 угловых секунд ниже и правее звезды (рис. 8).

Галактика SDSS J084818.23+045643.1 является рентгеновским источником 1RXS J084818.2+045709, и, очевидно, ей необходимо дать предпочтение при идентификации с объектом из каталога eFEDS. У галактики SDSS J090807.97-004613.9 ранее рентгеновское излучение не зафиксировано, но, учитывая её близость к положению объекта из eFEDS, возможно именно она яылется кандидатом на отождествление. Зелёные окружности на рисунках – область ошибок в определении координат по eFEDS, красная окружность – утроенный диаметр.

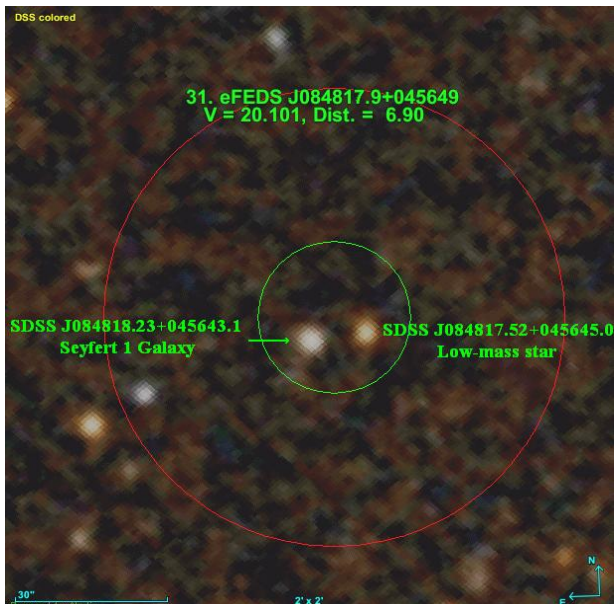

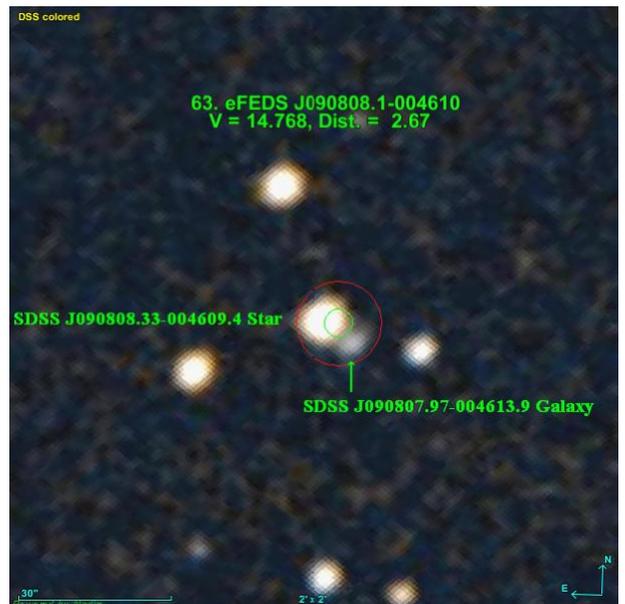

Рис. 7.                                                              Рис. 8.

Отметим, также, что в девять областей локализации eFEDS попадает более одного объекта, а 12 звёзд ранее были классифицированы, как рентгеновские источники. Более подробно информация об объектах приведена в конце таблицы с идентификацией источников.

## Заключение

Среди объектов глубокого обзора экваториальной области неба, выполненного телескопом eROSITA на обсерватории SRG, произведена идентификация звёзд, попадающих в 110 полей, ограниченных радиусом ошибок в определении рентгеновских координат. 12 звёзд ранее были классифицированы, как рентгеновские источники, что подтверждает правильность независимого отождествления. В областях идентификации двух звёзд были обнаружены галактики, одна из которых является известным рентгеновским объектом. Несколько отождествлений содержат близко расположенные объекты. В приложении 1 и 2 даны: список идентифицированных объектов с комментариями к нему и поисковые карты. Все области были проанализированы визуально.

Список объектов, ассоциированных с рентгеновскими источниками, зарегистрированными обсерваторий eROSITA, будет использован для анализа проявлений эруптивной активности звёзд по данным долговременных оптических наблюдений, выполняемых космическими и наземными средствами. А объекты, удовлетворяющие требованиям для включения в новый Каталог, будут присоединены к нему.

## Литература

1. Predehl P., Andritschke R., Arefiev V. et al. 2021, A&A, 647, A1.
2. Sunyaev R., Arefiev V., Babyshkin V. et al. 2021, arXiv e-prints, arXiv:2104.13267.
3. Brunner H., Liu T., Lamer G. et al. 2021, arXiv e-prints, arXiv:2106.14517.
4. Гершберг Р.Е., Клиорин Н.И., Пустильник Л.А., Шляпников А.А. "Физика звёзд средних и малых масс с активностью солнечного типа" / М.: ФИЗМАТЛИТ, 2020. – 768 с. – ISBN 978-5-9221-1881-1. Приложение. Доступ к каталогу: http://craocrimea.ru/~aas/CATALOGUEs/CSAST/CSAST.html
5. Доступ к каталогу eFEDS: https://erosita.mpe.mpg.de/edr/eROSITAObservations/Catalogues/
6. Gaia collaboration. Gaia data release 2 (Gaia DR2) // 2018A&A...616A...1G
7. Wenger et al. The SIMBAD astronomical database // A&AS, 2000, 143, 9.
8. Ochsenbein F., Bauer P., Marcout J. The VizieR database of astronomical catalogues // 2000A&AS..143...23O
9. Bonnarel F., Fernique P., Bienaymé O. The ALADIN interactive sky atlas. A reference tool for identification of astronomical sources // 2000A&AS..143...33B
10. SAO/NASA ADS. Доступ к рессурсу: https://ui.adsabs.harvard.edu/about/

# Search for red dwarfs among X-ray objects of the deep survey of the equatorial region of sky using eROSITA data
### Aleksey A. Shlyapnikov

## The list of identification objects

**Data description:**
1. № - record number.
2. RAJ_Corr - Right ascension (J2000), corrected (deg).
3. DEJ_Corr - Declination (J2000), corrected (deg).
4. ErCor - Combined positional error, corrected (arcsec).
5. eFEDS_Name – Name from eFEDS catalogue.
6. RAJ - Right ascension (J2000), for probable candidate (deg).
7. DEJ - Declination (J2000), for probable candidate (deg).
8. Dist – Distance betvin eFEDS sourreces and probable candidate (arcsec).
9. SIMBAD_Basic_Identification – SIMBAD basic identification.
10. Type - Object classification in SIMBAD.
11. V_mag – V magnitude from SIMBAD, or calculated from GAIA or SDSS.
12. SpTyp – spectral type from SIMBAD, SDSS or calculated from GAIA.
13. XF – detected X-ray flux.
14. Teff – Effective temperature (K).
15. Rsol – Radius in solars unit.
16. Lsol – Luminosity in solars unit.
17. Rotation – Rotation period from VSX (day).
18. Note – Comments.

| № | RAJ_Corr | DEJ_Corr | ErCor | eFEDS_Name | RAJ | DEJ | Dist | SIMBAD_Basic_Identification | Type | V_mag | SpTyp | XF | Teff | Rsol | Lsol | Rotation | Note |
|---|----------|----------|-------|------------|-----|-----|------|------------------------------|------|-------|-------|----|------|------|------|----------|------|
| 1 | 2 | 3 | 4 | 5 | 6 | 7 | 8 | 9 | 10 | 11 | 12 | 13 | 14 | 15 | 16 | 17 | 18 |
| 1 | 127.19843 | 0.02398 | 1.70 | eFEDS J082847.6+000126 | 127.19855 | 0.02426 | 0.95 | StKM 1-703 | Star | 11.112 | K5V | | 4669.20 | 0.67 | 0.190 | | |
| 2 | 128.10217 | 2.00864 | 4.43 | eFEDS J083224.5+020031 | 128.10194 | 2.01036 | 6.00 | TYC214-113-1 | PM* | 11.221 | K3V | | 4898.37 | 0.80 | 0.335 | | |
| 3 | 128.43294 | 4.23203 | 2.16 | eFEDS J083343.9+041355 | 128.43248 | 4.23275 | 2.47 | TYC218-1167-1 | Star | 11.580 | K3.5V | | 5100.33 | 0.57 | 0.181 | | |
| 4 | 128.63041 | -0.72618 | 0.98 | eFEDS J083431.3-004334 | 128.63188 | -0.72606 | 4.70 | HD72760 | PM* | 7.288 | K0V | X | 5339.00 | 0.83 | 0.501 | | |
| 5 | 128.70561 | 0.29243 | 6.36 | eFEDS J083449.3+001733 | 128.70598 | 0.29422 | 6.42 | TYC210-1860-1 | Star | 12.284 | G9V | | 5454.00 | 0.89 | 0.636 | | 1 |
| 6 | 128.84225 | 3.72338 | 3.97 | eFEDS J083522.1+034324 | 128.84286 | 3.72380 | 2.77 | TYC214-1435-1 | Star | 11.490 | K0V | | 5329.23 | 0.81 | 0.481 | | 2 |
| 7 | 128.87554 | -0.28369 | 4.71 | eFEDS J083530.1-001701 | 128.87637 | -0.28256 | 4.72 | TYC 4862-1374-1 | Star | 11.728 | K1V | | 5210.53 | 0.93 | 0.571 | | |
| 8 | 128.95150 | -2.25107 | 2.43 | eFEDS J083548.4-021504 | 128.95247 | -2.25191 | 3.85 | Gaia DR2 3072788382888799232 | Star | 13.165 | M1.5V | | 3848.00 | 0.59 | 0.070 | | |
| 9 | 129.21451 | 2.41350 | 1.44 | eFEDS J083651.5+022449 | 129.21483 | 2.41368 | 1.26 | TYC214-788-1 | Star | 11.649 | K3V | | 4836.33 | 0.78 | 0.297 | | |
| 10 | 129.23753 | 4.30749 | 3.50 | eFEDS J083657.0+041827 | 129.23711 | 4.30720 | 1.81 | TYC218-1127-1 | Star | 11.564 | K2V | | 5114.83 | 0.76 | 0.360 | | |
| 11 | 129.33933 | 2.98113 | 0.89 | eFEDS J083721.4+025852 | 129.33940 | 2.98119 | 0.46 | TYC214-1872-1 | Star | 11.312 | K5V | X | 4859.67 | 0.58 | 0.166 | 0.690566 | |
| 12 | 129.42353 | 2.11404 | 4.45 | eFEDS J083741.6+020651 | 129.42375 | 2.11400 | 0.89 | TYC215-961-1 | PM* | 9.779 | G4V | | 5654.33 | 0.94 | 0.821 | | |
| 13 | 129.53792 | 2.51159 | 1.61 | eFEDS J083809.1+023042 | 129.53847 | 2.51214 | 2.68 | BD+032024 | PM* | 9.635 | G9V | | 5405.00 | 0.84 | 0.541 | | |
| 14 | 129.55411 | -1.15606 | 2.51 | eFEDS J083813.0-010922 | 129.55521 | -1.15555 | 3.76 | TYC 4863-571-1 | PM* | 10.933 | K2.5V | | 4951.92 | 0.80 | 0.347 | | |
| 15 | 129.70090 | 2.58547 | 4.94 | eFEDS J083848.2+023453 | 129.70172 | 2.59207 | 5.73 | Gaia DR2 3079158506583841024 | Star | 12.816 | K2V | | 5033.00 | 0.86 | 0.426 | | |
| 16 | 129.92364 | -1.19269 | 5.18 | eFEDS J083941.7-011134 | 129.92474 | -1.19283 | 3.29 | TYC 4863-1195-1 | PM* | 11.438 | K4.5V | | 4644.50 | 0.68 | 0.193 | | |
| 17 | 130.11169 | 0.27347 | 4.84 | eFEDS J084026.8+001624 | 130.11285 | 0.27315 | 3.99 | TYC211-253-1 | PM* | 11.556 | K2V | | 5046.62 | 0.77 | 0.342 | | |
| 18 | 130.21775 | -2.25922 | 3.97 | eFEDS J084052.3-021533 | 130.21886 | -2.25984 | 3.70 | TYC 4867-1258-1 | Star | 11.532 | K1.5V | | 5161.80 | 0.81 | 0.416 | | 3 |

| 1 | 2 | 3 | 4 | 5 | 6 | 7 | 8 | 9 | 10 | 11 | 12 | 13 | 14 | 15 | 16 | 17 | 18 |
|---|---|---|---|---|---|---|---|---|---|---|---|---|---|---|---|---|---|
| 19 | 130.27402 | 2.85666 | 2.62 | eFEDS J084105.8+025124 | 130.27444 | 2.85717 | 2.28 | TYC215-1164-1 | PM* | 10.317 | K3V | | 4898.03 | 0.81 | 0.341 | | |
| 20 | 130.36591 | -1.23443 | 2.63 | eFEDS J084127.8-011404 | 130.36639 | -1.23407 | 1.67 | UCAC4 444-047525 | PM* | 14.402 | M0.5V | | 3837.00 | 0.65 | 0.083 | | |
| 21 | 130.51008 | 1.05766 | 4.24 | eFEDS J084202.4+010318 | 130.51021 | 1.05732 | 1.40 | TYC211-1314-1 | Star | 11.167 | G8V | | 5459.00 | 1.03 | 0.847 | | |
| 22 | 130.60779 | -0.29843 | 6.96 | eFEDS J084225.9-001754 | 130.60979 | -0.29900 | 6.99 | TYC 4863-498-1 | Star | 12.349 | K0V | | 5424.00 | 0.90 | 0.630 | | |
| 23 | 130.64573 | 1.41727 | 2.82 | eFEDS J084235.0+012502 | 130.64516 | 1.41761 | 2.42 | TYC211-689-1 | Star | 10.537 | K2.5V | | 4978.76 | 1.01 | 0.570 | | |
| 24 | 131.02537 | 0.76238 | 5.90 | eFEDS J084406.1+004545 | 131.02563 | 0.76407 | 6.00 | TYC211-796-1 | Star | 11.694 | K1V | | 5176.50 | 0.86 | 0.475 | | 4 |
| 25 | 131.17473 | 0.73792 | 0.94 | eFEDS J084441.9+004417 | 131.17555 | 0.73776 | 2.74 | TYC211-1502-1 | PM* | 10.259 | K4.5V | X | 4808.92 | 0.90 | 0.390 | 35.170158 | |
| 26 | 131.67697 | 3.49646 | 4.03 | eFEDS J084642.5+032947 | 131.67702 | 3.49627 | 1.15 | TYC216-895-1 | PM* | 10.610 | K2V | | 5031.60 | 0.98 | 0.557 | | |
| 27 | 131.73112 | 3.39503 | 2.29 | eFEDS J084655.5+032342 | 131.73207 | 3.39571 | 4.24 | TYC216-1010-1 | Star | 11.813 | K1.5V | | 5051.33 | 0.72 | 0.306 | | |
| 28 | 131.82035 | 3.01396 | 3.45 | eFEDS J084716.9+030050 | 131.82184 | 3.01439 | 5.69 | LSPM J0847+0300 | l-m* | 0.000 | | | - | - | - | | 5 |
| 29 | 131.86254 | 0.23254 | 1.61 | eFEDS J084727.0+001357 | 131.86316 | 0.23260 | 1.84 | TYC212-1457-1 | PM* | 11.336 | K2V | | 5040.50 | 0.82 | 0.392 | | |
| 30 | 131.89303 | 0.00000 | 2.28 | eFEDS J084734.3+000135 | 131.89319 | 0.02709 | 2.27 | BD+002393B | Star | 9.012 | G8V | | 5576.00 | 1.02 | 0.899 | | |
| *31 | 132.07471 | 4.94681 | 15.11 | eFEDS J084817.9+045649 | 132.07298 | 4.94583 | 6.90 | SDSS J084817.52+045645.0 | l-m* | 20.101 | M2.5V | | - | - | - | | 5, 6 |
| 32 | 132.29659 | 0.49321 | 1.52 | eFEDS J084911.2+002936 | 132.29761 | 0.49346 | 3.42 | BD+012178 | PM* | 9.951 | K1.5V | X | 5100.00 | 0.79 | 0.377 | | |
| 33 | 132.30193 | 3.48500 | 1.47 | eFEDS J084912.5+032000 | 132.30220 | 3.48476 | 1.77 | HD75302 | PM* | 7.435 | G6V | X | 5699.33 | 0.94 | 0.842 | | |
| 34 | 132.79900 | 5.16322 | 1.71 | eFEDS J085111.8+050948 | 132.79921 | 5.16309 | 1.72 | BD+052067 | PM* | 9.148 | G8V | | 5530.00 | 0.94 | 0.752 | | |
| 35 | 132.80496 | 1.33454 | 1.42 | eFEDS J085113.2+012004 | 132.80530 | 1.33481 | 1.32 | HD75639 | PM* | 9.011 | K2V | | 5024.10 | 0.71 | 0.293 | | |
| 36 | 132.81335 | -2.06289 | 8.08 | eFEDS J085115.2+023004 | 132.81340 | -2.06288 | 0.74 | TYC 4868-1117-1 | Star | 11.792 | K6.5V | | 4185.00 | 0.78 | 0.169 | | 7 |
| 37 | 132.83294 | 3.46059 | 3.66 | eFEDS J085119.9+032738 | 132.83348 | 3.46194 | 5.01 | TYC216-289-1 | PM* | 10.669 | K2.5V | | 4915.29 | 0.78 | 0.324 | | |
| 38 | 132.97546 | 5.81898 | 7.19 | eFEDS J085154.1+054908 | 132.97534 | 5.81820 | 3.38 | SDSS J085154.07+054905.4 | Star | 14.956 | K0V | | 5422.67 | 0.80 | 0.500 | | |
| 39 | 132.98434 | -0.77111 | 1.49 | eFEDS J085156.2-004616 | 132.98486 | -0.77050 | 2.24 | TYC 4864-260-1 | Star | 11.881 | G7V | | 5803.67 | 0.79 | 0.641 | | |
| 40 | 133.04400 | 1.45513 | 4.22 | eFEDS J085210.6+012718 | 133.04358 | 1.45643 | 4.77 | TYC212-329-1 | Star | 11.388 | K3V | | 4860.35 | 0.90 | 0.407 | | 8 |
| 41 | 133.29281 | -0.98661 | 4.00 | eFEDS J085310.3-005912 | 133.29318 | -0.98700 | 1.52 | TYC 4865-809-1 | Star | 11.510 | G4V | | 5752.00 | 0.98 | 0.938 | | |
| 42 | 133.34866 | 1.88110 | 7.45 | eFEDS J085323.7+015252 | 133.34972 | 1.88351 | 4.29 | SDSS J085323.93+015254.3 | l-m* | 20.122 | M5.5V | | - | - | - | | 5 |
| 43 | 133.42820 | 4.48503 | 4.10 | eFEDS J085342.8+042906 | 133.42946 | 4.48423 | 6.05 | SDSS J085343.07+042903.2 | Star | 18.155 | M5.5V | | 4067.17 | 0.00 | 0.000 | | |
| 44 | 133.79566 | 1.27338 | 2.80 | eFEDS J085511.0+011624 | 133.79553 | 1.27204 | 5.06 | TYC213-1211-1 | PM* | 11.930 | K1.5V | | 5164.00 | 0.87 | 0.486 | | 9 |
| 45 | 134.14459 | 2.52806 | 3.57 | eFEDS J085634.7+023141 | 134.14420 | 2.52759 | 2.54 | 2MASS J08563459+0231397 | Star | 12.187 | K3.5V | | 4811.28 | 1.03 | 0.507 | | |
| 46 | 134.27526 | -1.30452 | 9.11 | eFEDS J085706.1-011816 | 134.27431 | -1.30338 | 5.87 | TYC 4865-550-1 | PM* | 12.013 | G9V | | 5446.33 | 0.98 | 0.755 | | |
| 47 | 134.43835 | 2.82549 | 5.23 | eFEDS J085745.2+024932 | 134.43692 | 2.82546 | 5.00 | TYC217-1083-1 | Star | 11.300 | G0V | | 5897.50 | 0.96 | 1.000 | | |
| 48 | 134.51207 | -1.52615 | 1.91 | eFEDS J085802.9-013134 | 134.51193 | -1.52601 | 1.42 | TYC 4865-710-1 | Star | 11.484 | K0.5V | | 5254.58 | 1.01 | 0.703 | | |
| 49 | 134.79389 | 3.22182 | 1.55 | eFEDS J085910.5+031319 | 134.79434 | 3.22111 | 3.44 | 2MASS J08591064+0313160 | PM* | 13.188 | M0V | X | 3958.67 | 0.98 | 0.214 | | |
| 50 | 134.95820 | -1.57913 | 6.47 | eFEDS J085950.0-013445 | 134.95921 | -1.58060 | 5.93 | BD-012173B | PM* | 10.252 | K2V | | 5057.43 | 0.76 | 0.338 | | |
| 51 | 135.30278 | 2.74584 | 3.75 | eFEDS J090112.7+024445 | 135.30329 | 2.74522 | 2.70 | HD72193B | PM* | 9.397 | G7V | | 5628.00 | 1.01 | 0.931 | | |
| 52 | 135.53896 | 4.12879 | 2.94 | eFEDS J090209.4+040744 | 135.53957 | 4.12857 | 2.90 | SDSS J090209.49+040742.8 | l-m* | 17.771 | M4.5V | | 4613.21 | 0.00 | 0.000 | | 5, 10 |
| 53 | 135.54426 | 0.92304 | 3.35 | eFEDS J090210.6+005523 | 135.54445 | 0.92246 | 2.28 | TYC226-1775-1 | Star | 11.253 | G4V | | 5709.81 | 0.95 | 0.871 | | |
| 54 | 135.63181 | -0.87958 | 2.31 | eFEDS J090231.6-005246 | 135.63165 | -0.87926 | 1.69 | TYC 4878-1191-1 | Star | 11.460 | K3.5V | | 5033.25 | 0.61 | 0.218 | | |
| 55 | 135.69856 | 0.67308 | 2.44 | eFEDS J090247.7+004023 | 135.69833 | 0.67338 | 1.49 | TYC226-2364-1 | Star | 11.503 | G4V | | 5732.00 | 1.01 | 1.100 | | |
| 56 | 135.73150 | 2.22290 | 3.60 | eFEDS J090255.6+021322 | 135.73109 | 2.22425 | 4.82 | TYC226-1529-1 | Star | 11.609 | G9V | | 5540.00 | 0.90 | 0.691 | | |
| 57 | 135.82111 | 2.61683 | 2.77 | eFEDS J090317.1+023701 | 135.82127 | 2.61664 | 1.15 | CCDM J09033+0237B | PM* | 12.263 | K6.5V | | 4433.50 | 0.60 | 0.124 | | |
| 58 | 136.01421 | -0.38688 | 5.21 | eFEDS J090403.4-002313 | 136.01466 | -0.38630 | 2.34 | TYC 4878-293-1 | Star | 11.944 | G9V | | 5413.25 | 0.86 | 0.574 | | |
| 59 | 136.09356 | 0.26729 | 3.23 | eFEDS J090422.5+001602 | 136.09473 | 0.26750 | 3.88 | SDSS J090422.74+001603.0 | l-m* | 0.000 | | | - | - | - | | 5 |
| 60 | 136.12252 | 2.58694 | 5.57 | eFEDS J090429.4+023513 | 136.12323 | 2.58631 | 3.69 | SDSS J090429.58+023510.7 | l-m* | 20.220 | M5.5V | | - | - | - | | 5 |

| 1 | 2 | 3 | 4 | 5 | 6 | 7 | 8 | 9 | 10 | 11 | 12 | 13 | 14 | 15 | 16 | 17 | 18 |
|---|---|---|---|---|---|---|---|---|----|----|----|----|----|----|----|----|----|
| 61 | 136.23266 | 3.88549 | 1.05 | eFEDS J090455.8+035308 | 136.23282 | 3.88534 | 1.34 | TYC229-1396-1 | Star | 11.312 | G9V | X | 5409.50 | 0.89 | 0.607 | | |
| 62 | 136.41202 | 3.92028 | 4.57 | eFEDS J090538.9+035513 | 136.41132 | 3.92068 | 2.36 | TYC229-1438-1 | PM* | 11.341 | K1.5V | | 5166.00 | 0.81 | 0.424 | | |
| *63 | 137.03382 | -0.76932 | 2.84 | eFEDS J090808.1-004610 | 137.03473 | -0.76928 | 2.67 | SDSS J090808.33-004609.4 | Star | 14.768 | K4.5V | | 4946.75 | 0.86 | 0.397 | | 11 |
| 64 | 137.06602 | 4.37278 | 3.03 | eFEDS J090815.8+042222 | 137.06659 | 4.37303 | 2.58 | TYC229-247-1 | Star | 11.746 | G9V | | 5545.90 | 0.86 | 0.623 | | |
| 65 | 137.11799 | 0.90175 | 4.51 | eFEDS J090828.3+005406 | 137.11894 | 0.90186 | 3.18 | StKM 1-750 | PM* | 10.735 | K5.5V | | 4447.00 | 0.69 | 0.169 | | |
| 66 | 137.37474 | -0.39895 | 4.70 | eFEDS J090929.9+002354 | 137.37569 | -0.39843 | 3.41 | HD78663 | PM* | 8.601 | K0V | | 5395.75 | 0.73 | 0.409 | | |
| 67 | 137.41923 | 1.74823 | 1.77 | eFEDS J090940.6+014454 | 137.41872 | 1.74885 | 2.77 | 2MASS J09094049+0144559 | Star | 14.862 | M3V | X | 3965.05 | 0.51 | 0.058 | | |
| 68 | 137.47465 | 5.20332 | 1.30 | eFEDS J090953.9+051212 | 137.47551 | 5.20349 | 3.76 | HD78727 | PM* | 8.334 | K3V | | 4876.11 | 0.82 | 0.344 | | |
| 69 | 137.55231 | 1.49411 | 1.95 | eFEDS J091012.6+012939 | 137.55133 | 1.49432 | 3.70 | BD+022151 | PM* | 10.199 | K2V | | 5106.00 | 0.81 | 0.403 | | |
| 70 | 137.59826 | 1.81027 | 4.74 | eFEDS J091023.6+014837 | 137.59747 | 1.81066 | 3.12 | TYC227-1263-1 | Star | 12.268 | G7V | | 5729.96 | 0.95 | 0.871 | | |
| 71 | 137.74183 | 3.60319 | 1.95 | eFEDS J091058.0+033611 | 137.74239 | 3.60361 | 2.59 | TYC230-1211-1 | Star | 10.846 | G9V | | 5527.00 | 0.95 | 0.766 | | |
| 72 | 137.79145 | 5.06317 | 2.15 | eFEDS J091109.9+050347 | 137.79158 | 5.06390 | 2.38 | TYC233-1643-1 | PM* | 10.505 | K1V | | 5197.95 | 0.80 | 0.419 | | |
| 73 | 137.98746 | -2.27653 | 6.15 | eFEDS J091157.0-021636 | 137.98781 | -2.27575 | 2.97 | TYC 4883-689-1 | Star | 11.121 | G9V | | 5371.00 | 0.87 | 0.562 | | |
| 74 | 138.14529 | -1.15867 | 3.50 | eFEDS J091234.9-010931 | 138.14607 | -1.15885 | 2.21 | TYC 4879-879-1 | Star | 11.560 | K2V | | 4997.12 | 0.77 | 0.334 | | |
| 75 | 138.16902 | 1.13897 | 3.59 | eFEDS J091240.6+010820 | 138.17036 | 1.13868 | 4.77 | TYC227-448-1 | PM* | 11.351 | K3V | | 4860.13 | 0.72 | 0.261 | | |
| 76 | 138.24373 | 0.51509 | 4.17 | eFEDS J091258.5+003054 | 138.24356 | 0.51560 | 1.98 | TYC227-1961-1 | Star | 11.221 | K1.5V | | 5169.24 | 0.78 | 0.391 | | |
| 77 | 138.37700 | -2.26637 | 1.83 | eFEDS J091330.5-021559 | 138.37761 | -2.26621 | 1.42 | TYC 4883-646-1 | Star | 11.488 | G7V | | 5500.86 | 1.03 | 0.882 | | |
| 78 | 138.42875 | 5.51785 | 2.17 | eFEDS J091342.9+053104 | 138.42888 | 5.51817 | 1.36 | TYC233-922-1 | Star | 11.697 | K2V | | 5086.76 | 0.85 | 0.432 | | |
| 79 | 138.44682 | 4.41977 | 1.82 | eFEDS J091347.2+042511 | 138.44586 | 4.42085 | 4.51 | HD79375 | PM* | 8.839 | G9V | | 5456.00 | 0.84 | 0.562 | | |
| 80 | 138.51276 | -1.96814 | 5.13 | eFEDS J091403.1-015805 | 138.51317 | -1.96867 | 1.91 | UCAC3 177-111583 | Star | 11.895 | K6V | | 4325.72 | 0.79 | 0.196 | | |
| 81 | 138.53781 | 0.25636 | 4.00 | eFEDS J091409.1+001523 | 138.53806 | 0.25621 | 0.80 | G 114-50 | PM* | 9.740 | K1.5V | | 5087.13 | 0.85 | 0.441 | | |
| 82 | 138.72288 | 4.44338 | 0.81 | eFEDS J091453.5+042636 | 138.72354 | 4.44284 | 3.75 | HD79555 | SB* | 7.961 | K5V | X | 4861.00 | 0.69 | 0.238 | | |
| 83 | 138.73773 | 0.58091 | 4.00 | eFEDS J091457.1+003451 | 138.73668 | 0.58016 | 4.98 | TYC227-1052-1 | Star | 11.544 | K2.5V | | 4981.67 | 0.82 | 0.375 | | |
| 84 | 138.99508 | 0.99016 | 9.70 | eFEDS J091558.8+005925 | 138.99557 | 0.99006 | 1.88 | TYC227-1250-1 | Star | 11.757 | K2V | | 5082.63 | 1.07 | 0.682 | | |
| 85 | 139.08505 | -2.03021 | 3.11 | eFEDS J091620.4-020149 | 139.08685 | -2.03000 | 5.66 | TYC 4884-117-1 | Star | 11.469 | G9V | | 5540.00 | 0.82 | 0.573 | | |
| 86 | 139.47407 | -0.06767 | 3.10 | eFEDS J091753.8-000404 | 139.47310 | -0.06844 | 4.85 | TYC 4880-182-1 | Star | 11.964 | K1V | | 5267.00 | 0.90 | 0.558 | | 12 |
| 87 | 139.59769 | 4.40708 | 4.78 | eFEDS J091823.4+042425 | 139.59695 | 4.40737 | 3.55 | BD+052155 | Star | 10.978 | G8V | | 5756.25 | 0.96 | 0.912 | | |
| 88 | 139.71186 | 2.75076 | 2.38 | eFEDS J091850.8+024503 | 139.71198 | 2.75086 | 0.58 | StKM 1-766 | Pec* | 11.719 | K7V | | 4193.92 | 0.88 | 0.217 | | |
| 89 | 140.33438 | -0.89878 | 3.33 | eFEDS J092120.3-005356 | 140.33369 | -0.89854 | 3.25 | TYC 4880-129-1 | Star | 10.279 | G4V | | 5740.05 | 0.96 | 0.900 | | |
| 90 | 140.42737 | 2.51755 | 0.96 | eFEDS J092142.6+023103 | 140.42783 | 2.51782 | 1.89 | TYC231-1587-1 | RoV* | 11.381 | K0V | X | 5382.25 | 0.97 | 0.707 | 0.53298 | 13 |
| 91 | 140.46699 | -0.99475 | 4.79 | eFEDS J092152.1-005941 | 140.46779 | -0.99317 | 6.13 | TYC 4880-510-1 | Star | 11.955 | G7V | | 5762.00 | 0.93 | 0.855 | | 14 |
| 92 | 140.63685 | 2.28481 | 3.37 | eFEDS J092232.8+021705 | 140.63707 | 2.28455 | 1.45 | BD+022191 | Star | 10.673 | G4V | | 5819.12 | 0.96 | 0.946 | | |
| 93 | 140.66924 | 4.63607 | 2.50 | eFEDS J092240.6+043810 | 140.67017 | 4.63616 | 3.94 | TYC231-751-1 | Star | 10.768 | G4V | | 5564.00 | 1.07 | 0.983 | | |
| 94 | 140.79596 | -0.03069 | 6.11 | eFEDS J092311.0-000150 | 140.79576 | -0.03115 | 2.10 | SDSS J092310.98-000152.0 | l-m* | 21.701 | M6.5V | | - | - | - | | 5 |
| 95 | 140.91793 | -1.33366 | 5.26 | eFEDS J092340.3-012001 | 140.91832 | -1.33399 | 1.33 | TYC 4881-1044-1 | PM* | 11.609 | K2V | | 5095.29 | 0.77 | 0.363 | | |
| 96 | 141.14889 | -0.47679 | 0.99 | eFEDS J092435.7-002836 | 141.14922 | -0.47717 | 1.51 | HD81266 | PM* | 9.400 | G4V | X | 5817.67 | 1.02 | 1.081 | | |
| 97 | 141.16123 | 2.34784 | 3.39 | eFEDS J092438.7+022052 | 141.16040 | 2.34798 | 2.96 | 2MASS J09243849+0220526 | Star | 14.695 | K2.5V | X | 4965.00 | 0.95 | 0.489 | | |
| 98 | 141.22633 | -1.25672 | 4.47 | eFEDS J092454.3-011524 | 141.22743 | -1.25624 | 3.67 | TYC 4881-943-1 | PM* | 11.799 | G7V | | 5891.45 | 0.98 | 1.048 | | |
| 99 | 141.67616 | 2.77034 | 5.84 | eFEDS J092642.3+024613 | 141.67560 | 2.77172 | 1.91 | TYC231-1177-1 | Star | 12.281 | K3V | | 4916.66 | 0.71 | 0.268 | | |
| 100 | 141.96313 | -0.96712 | 5.93 | eFEDS J092751.2-005802 | 141.96477 | -0.96728 | 5.27 | TYC 4881-1006-1 | Star | 11.356 | G9V | | 5422.16 | 0.85 | 0.565 | | |
| 101 | 142.27322 | -1.54818 | 4.45 | eFEDS J092905.6-013253 | 142.27399 | -1.54886 | 3.12 | BD-002202 | PM* | 10.715 | G5V | | 5669.83 | 0.97 | 0.883 | | |
| 102 | 142.60145 | 0.81437 | 3.45 | eFEDS J093024.3+004852 | 142.60139 | 0.81520 | 2.89 | TYC235-455-1 | Star | 10.961 | K2V | | 4982.94 | 0.70 | 0.275 | | |

| 1 | 2 | 3 | 4 | 5 | 6 | 7 | 8 | 9 | 10 | 11 | 12 | 13 | 14 | 15 | 16 | 17 | 18 |
|---|---|---|---|---|---|---|---|---|---|---|---|---|---|---|---|---|---|
| 103 | 142.71303 | 3.03842 | 11.91 | eFEDS J093051.1+030218 | 142.71223 | 3.03732 | 5.05 | LP547-74 | PM* | 11.052 | K4V | | 4814.42 | 0.66 | 0.212 | | |
| 104 | 143.46284 | 0.44298 | 4.18 | eFEDS J093351.1+002635 | 143.46304 | 0.44302 | 0.38 | BD+012323 | PM* | 10.654 | G7V | | 5547.25 | 0.98 | 0.815 | | |
| 105 | 143.50388 | -2.34120 | 4.98 | eFEDS J093400.9-022028 | 143.50464 | -2.34011 | 4.41 | TYC 4894-2501-1 | Star | 11.153 | K2V | | 4917.87 | 0.75 | 0.299 | | |
| 106 | 143.73632 | 2.57566 | 2.46 | eFEDS J093456.7+023432 | 143.73572 | 2.57609 | 2.40 | TYC238-487-1 | Star | 10.935 | G8V | | 5444.33 | 0.96 | 0.733 | | |
| 107 | 143.99680 | 5.02381 | 5.61 | eFEDS J093559.2+050126 | 143.99635 | 5.02401 | 1.01 | BD+052203 | PM* | 10.423 | K1.5V | | 5104.23 | 0.76 | 0.356 | | |
| 108 | 144.69487 | 4.80672 | 5.56 | eFEDS J093846.8+044824 | 144.69571 | 4.80749 | 4.27 | TYC238-1001-1 | Star | 11.221 | G7V | | 5598.95 | 0.93 | 0.768 | | |
| 109 | 145.04514 | 4.40373 | 3.34 | eFEDS J094010.8+042413 | 145.04672 | 4.40333 | 6.43 | TYC239-494-1 | Star | 12.207 | K2V | | 5020.78 | 0.75 | 0.322 | | |
| 110 | 145.14855 | 4.12192 | 1.89 | eFEDS J094035.7+040719 | 145.14870 | 4.12182 | 1.23 | TYC239-2160-1 | Star | 12.262 | K4V | | 5183.50 | 1.10 | 0.780 | | |

Note:
- The * marker next to the record number. Object is probably not a valid identification with red dwarf.
1. Two stars in the field: TYC 210-1860-1 and BD+00 2334 (V=9.45, Sp=K0).
2. Faint star near: Gaia DR2 3080242208438978688 (G_mag=18.058).
3. Two faint star near: Gaia DR1: Source ID 3072407707052266752 (top right, G_mag=16.220) and Source ID 3072407707054922112 (below a bright star, G_mag=19.879).
4. Faint star near. Below the bright star on the right (Gaia DR1: Source ID 3075322168782736384, G_mag=16.838).
5. l-m* - low mass star (low-mass*).
6. Two faint objects in error box: low mass star (SDSS J084817.52+045645.0, G_mag=19.211) and Seyfert 1 Galaxy (SDSS J084818.23+045643.1, G_mag=20.306) - probable identification.
7. Peculiar star near (UCAC4 440-047757, G_mag=13.839).
8. Object below the bright star (Gaia DR1: Source ID 3075652606386315520, G_mag=18.090).
9. Two bright stars in error box: TYC213-1211-1 and Gaia DR2 source ID 576837717289109248 (G_mag=10.593, Teff=5666.60, Rsol=1.18, Lsol=1.297).
10. Close pair of stars: SDSS J090209.49+040742.8 and Gaia DR2 source ID 578617478723879424 (V_mag=17.853, Sp=4.5V, Teff=4305.50).
11. The faint galaxy near (SDSS J090807.97-004613.9, V_mag=19.039). Below the bright star on the right (distance ~ 5 arcsec).
12. Two stars in the field: TYC 4880-182-1 and Gaia DR2 source ID 3842362856369154048 (G_mag=12.866, Teff=4789.52, Rsol=0.70, Lsol=0.233).
13. RoV* - rotation variable (RotV*).
14. Faint star near. Below the bright star on the right (Gaia DR2: Source ID 3839195777550937984, G_mag=18.560).

# Search for red dwarfs among X-ray objects of the deep survey of the equatorial region of sky using eROSITA data

Aleksey A. Shlyapnikov

## The identification charts

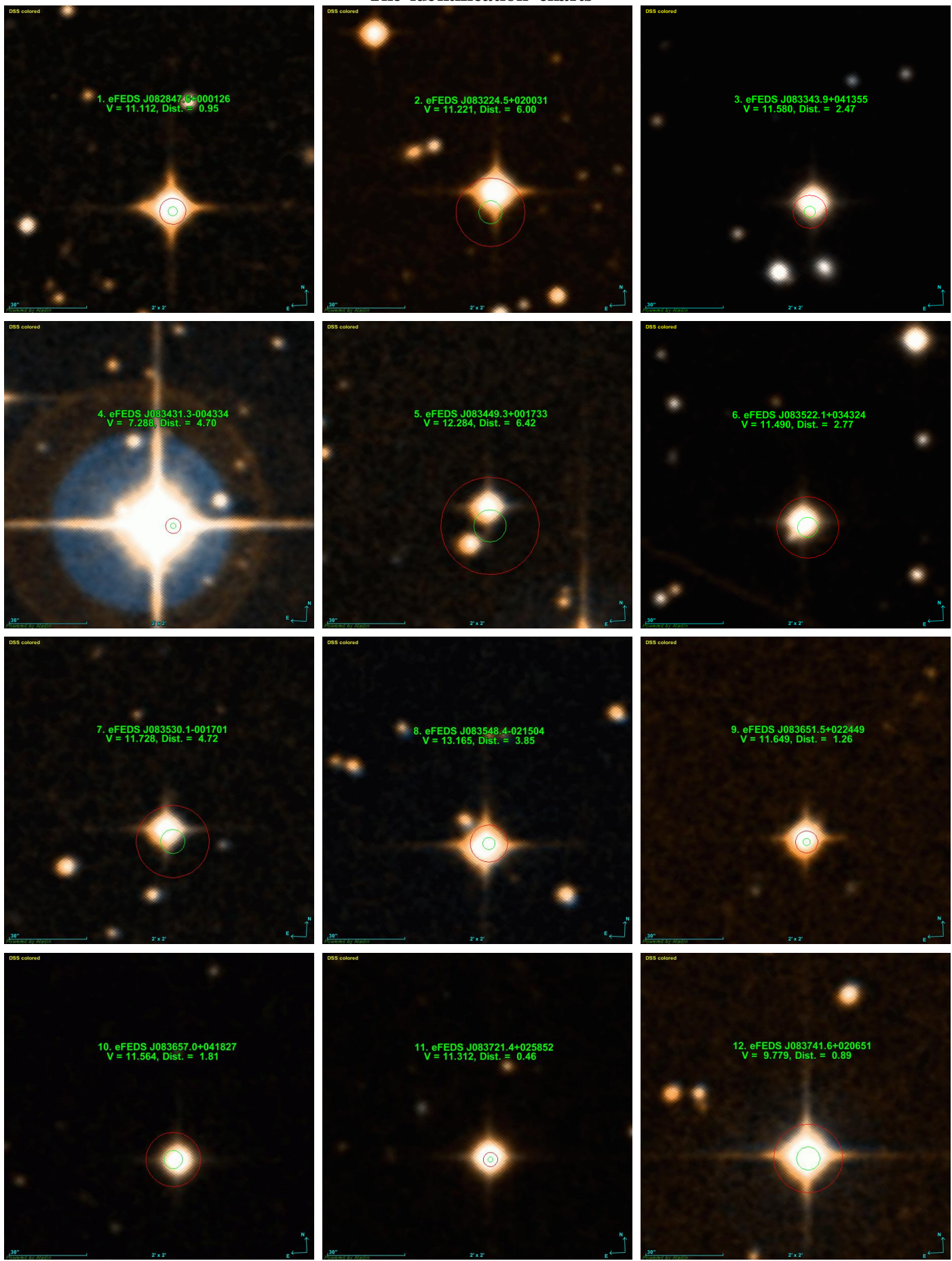

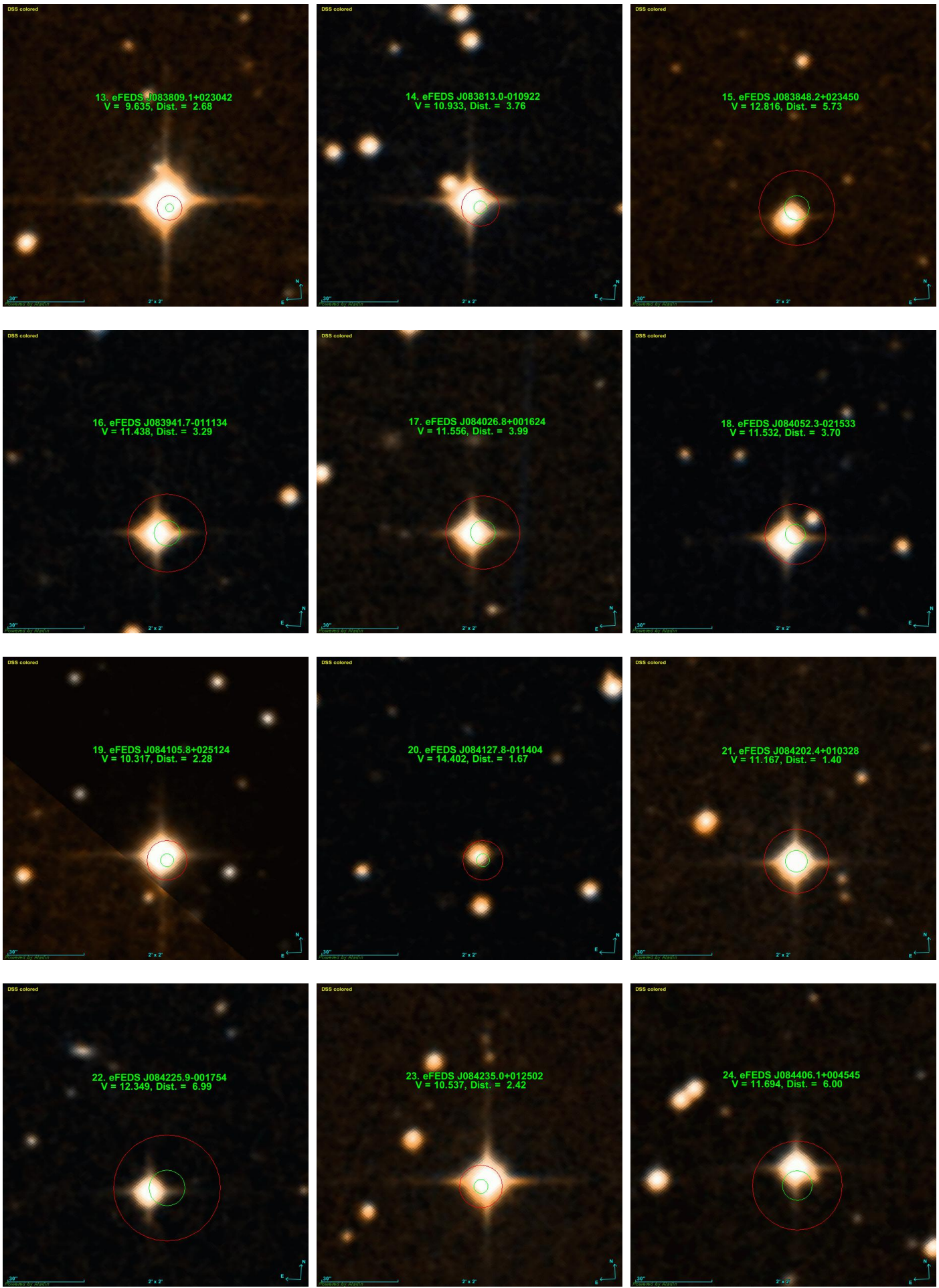

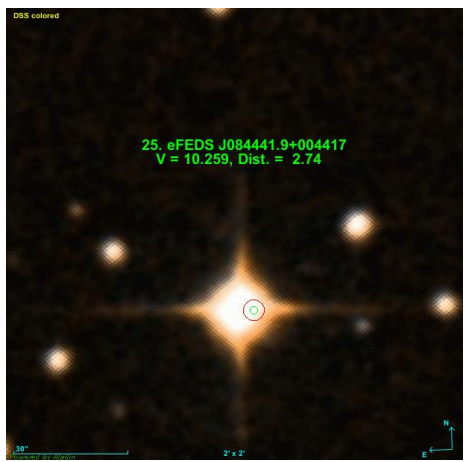
25. eFEDS J084441.9+004417
V = 10.259, Dist. = 2.74

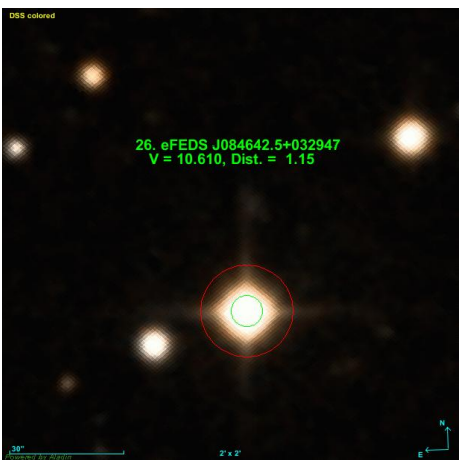
26. eFEDS J084642.5+032947
V = 10.610, Dist. = 1.15

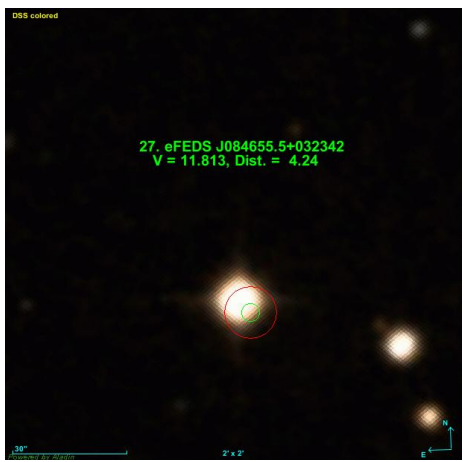
27. eFEDS J084655.5+032342
V = 11.813, Dist. = 4.24

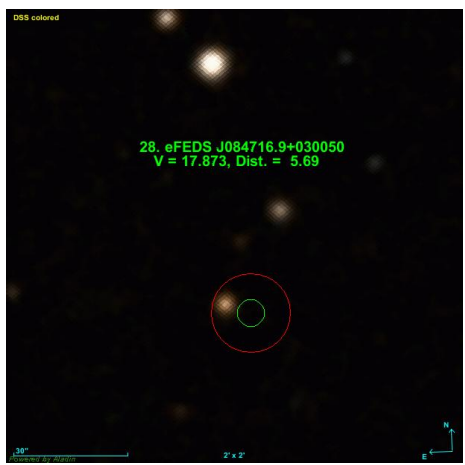
28. eFEDS J084716.9+030050
V = 17.873, Dist. = 5.69

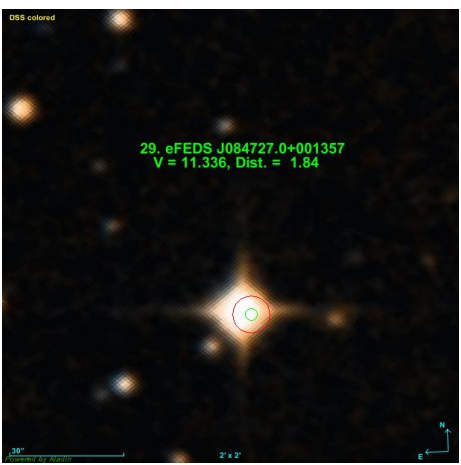
29. eFEDS J084727.0+001357
V = 11.336, Dist. = 1.84

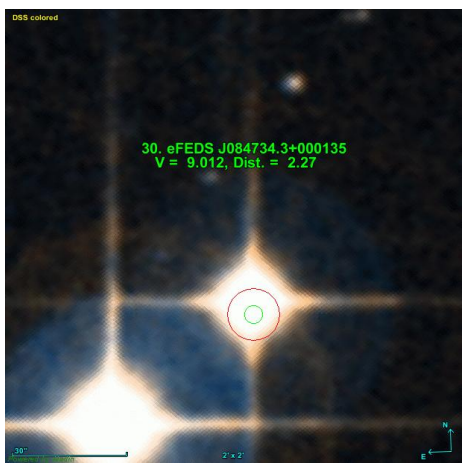
30. eFEDS J084734.3+000135
V = 9.012, Dist. = 2.27

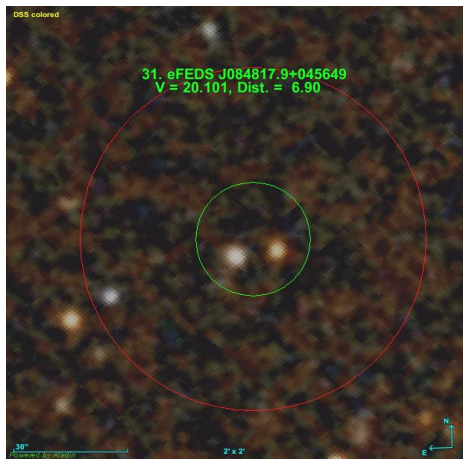
31. eFEDS J084817.9+045649
V = 20.101, Dist. = 6.90

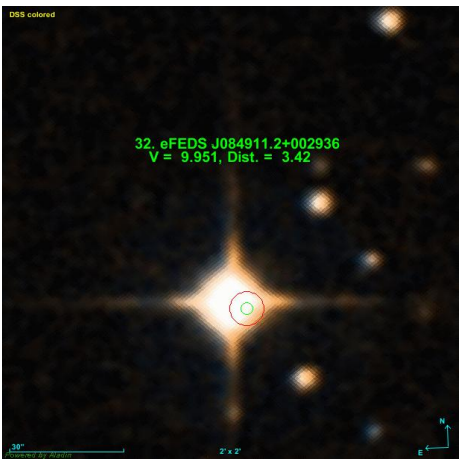
32. eFEDS J084911.2+002936
V = 9.951, Dist. = 3.42

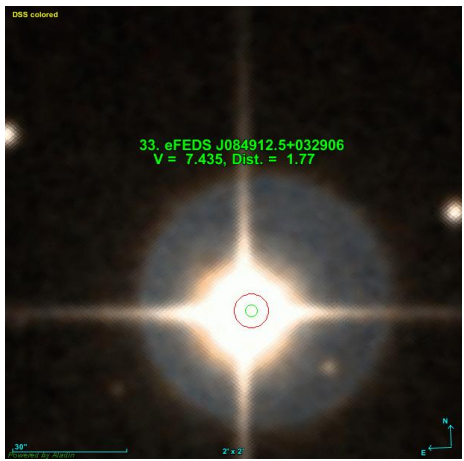
33. eFEDS J084912.5+032906
V = 7.435, Dist. = 1.77

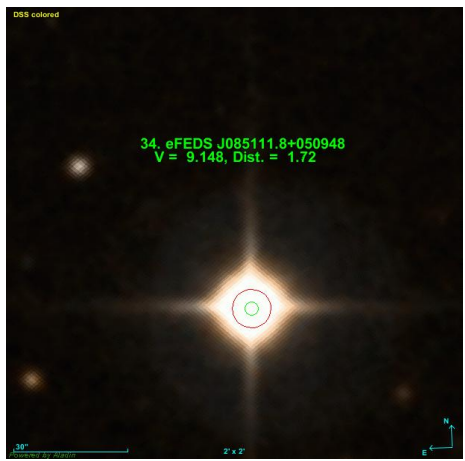
34. eFEDS J085111.8+050948
V = 9.148, Dist. = 1.72

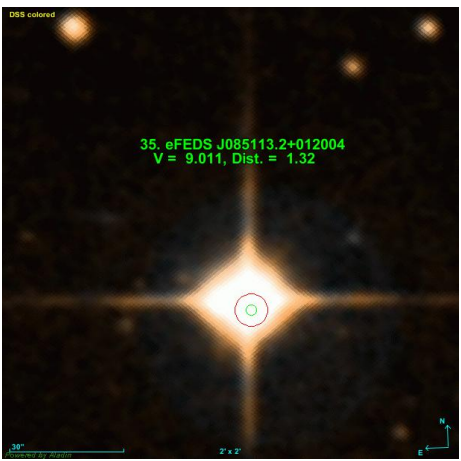
35. eFEDS J085113.2+012004
V = 9.011, Dist. = 1.32

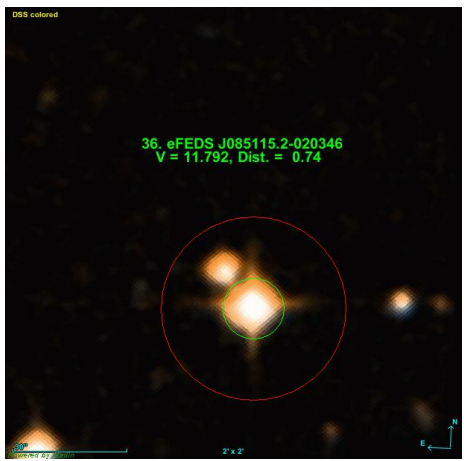
36. eFEDS J085115.2-020346
V = 11.792, Dist. = 0.74

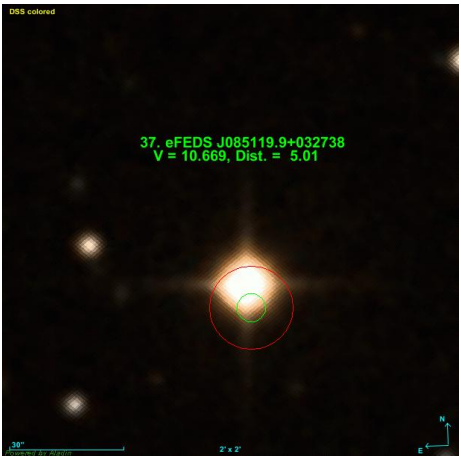

37. eFEDS J085119.9+032738
V = 10.669, Dist. = 5.01

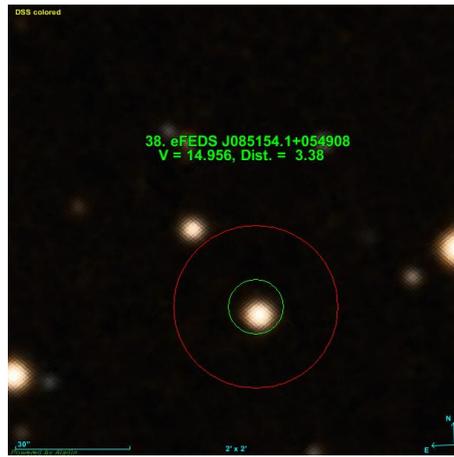

38. eFEDS J085154.1+054908
V = 14.956, Dist. = 3.38

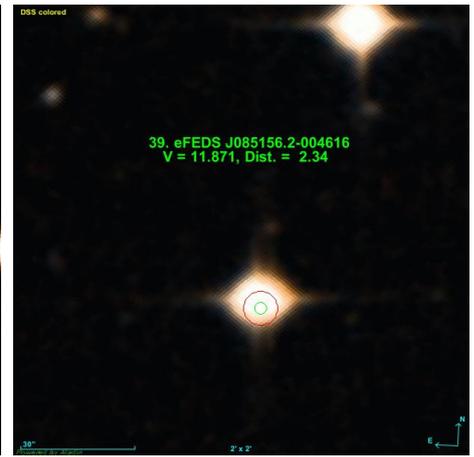

39. eFEDS J085156.2-004616
V = 11.871, Dist. = 2.34

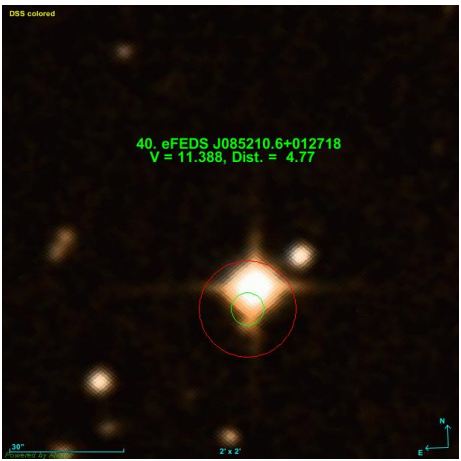

40. eFEDS J085210.6+012718
V = 11.388, Dist. = 4.77

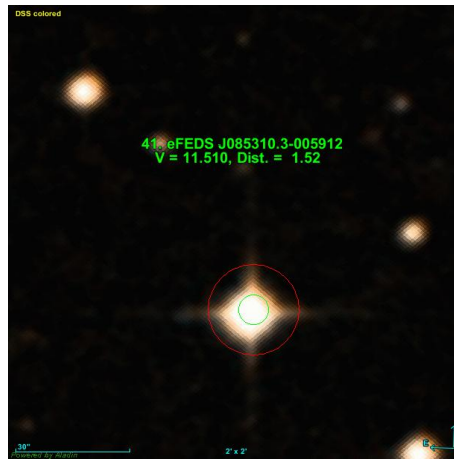

41. eFEDS J085310.3-005912
V = 11.510, Dist. = 1.52

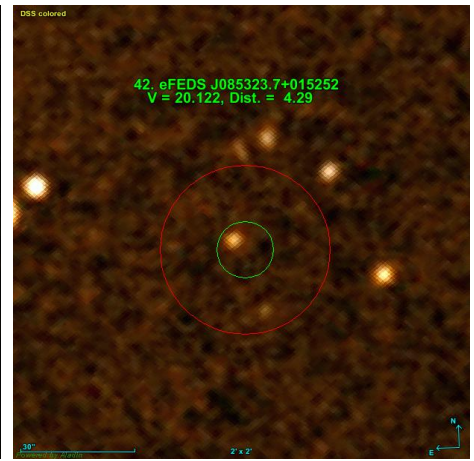

42. eFEDS J085323.7+015252
V = 20.122, Dist. = 4.29

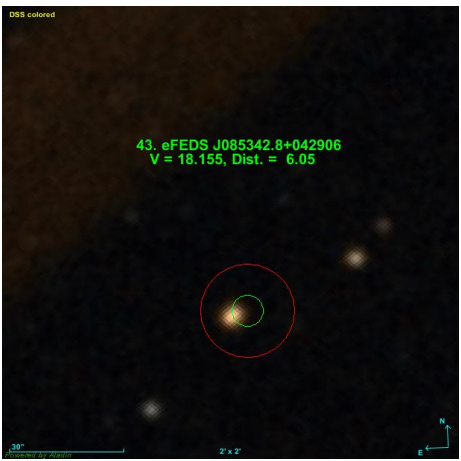

43. eFEDS J085342.8+042906
V = 18.155, Dist. = 6.05

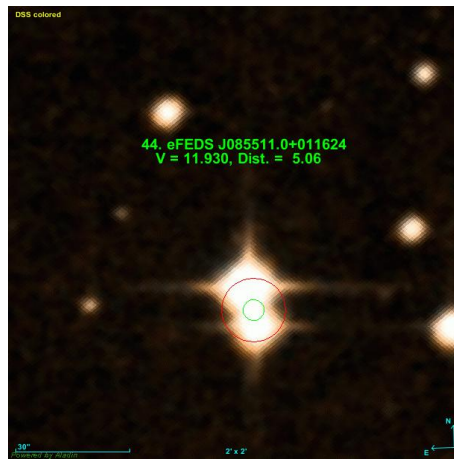

44. eFEDS J085511.0+011624
V = 11.930, Dist. = 5.06

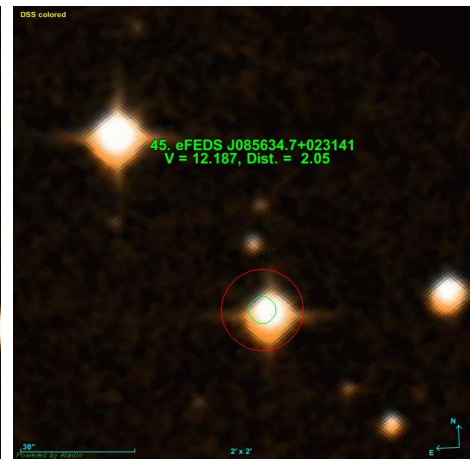

45. eFEDS J085634.7+023141
V = 12.187, Dist. = 2.05

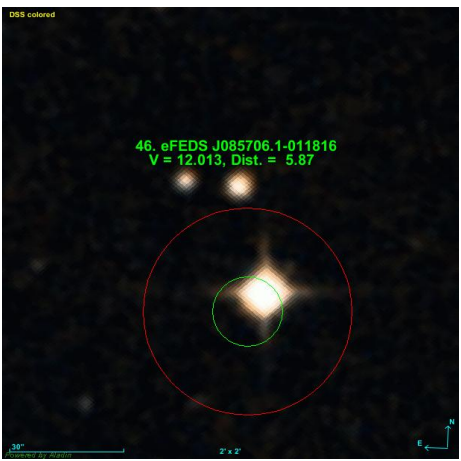

46. eFEDS J085706.1-011816
V = 12.013, Dist. = 5.87

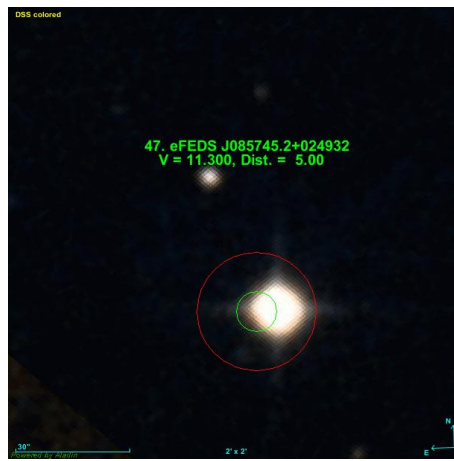

47. eFEDS J085745.2+024932
V = 11.300, Dist. = 5.00

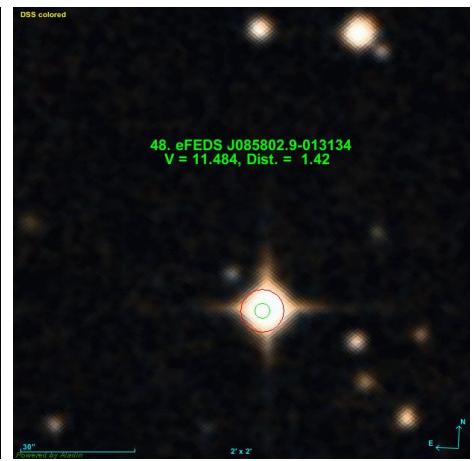

48. eFEDS J085802.9-013134
V = 11.484, Dist. = 1.42

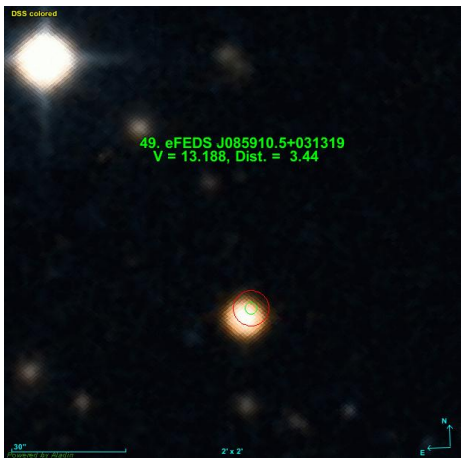
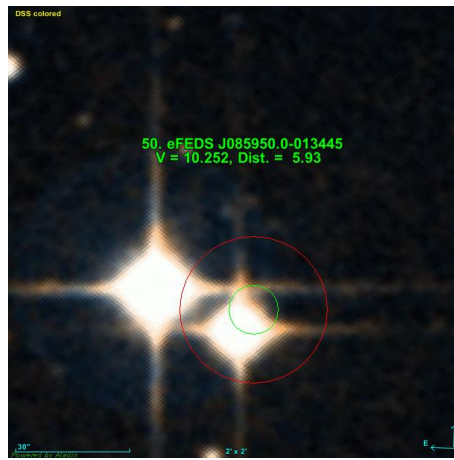
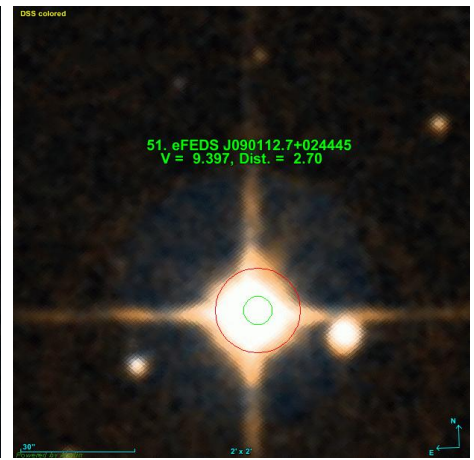

49. eFEDS J085910.5+031319
V = 13.188, Dist. = 3.44

50. eFEDS J085950.0-013445
V = 10.252, Dist. = 5.93

51. eFEDS J090112.7+024445
V = 9.397, Dist. = 2.70

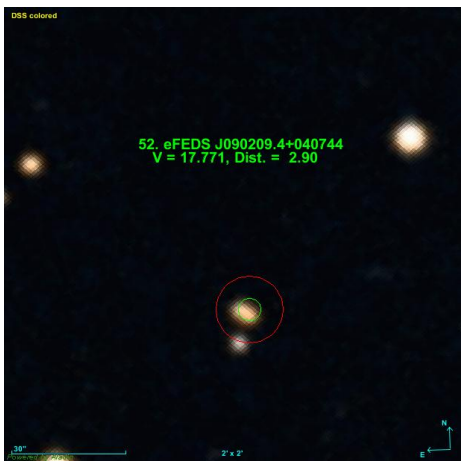
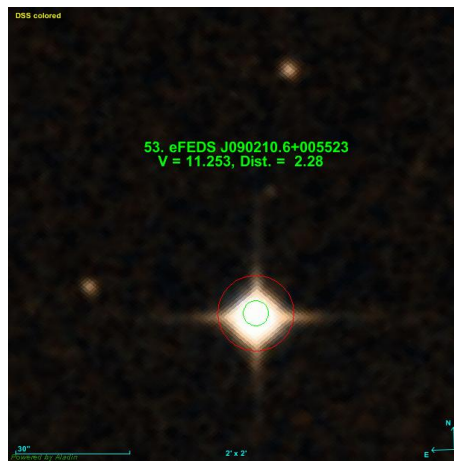
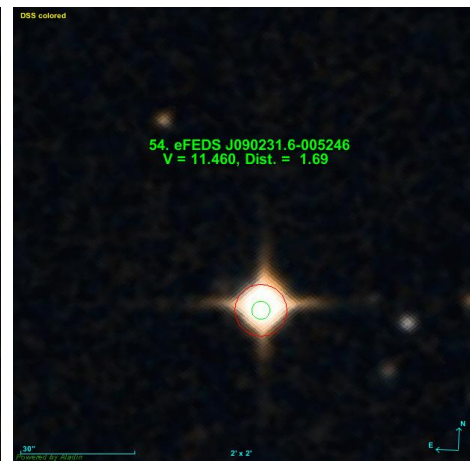

52. eFEDS J090209.4+040744
V = 17.771, Dist. = 2.90

53. eFEDS J090210.6+005523
V = 11.253, Dist. = 2.28

54. eFEDS J090231.6-005246
V = 11.460, Dist. = 1.69

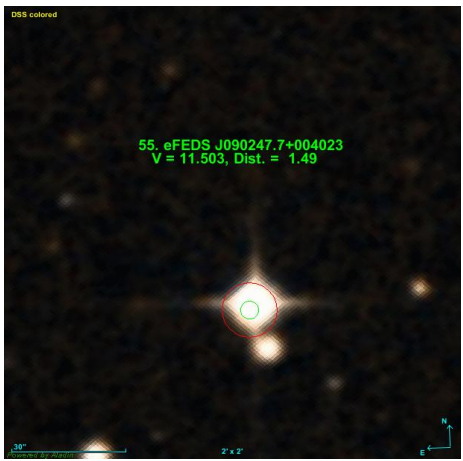
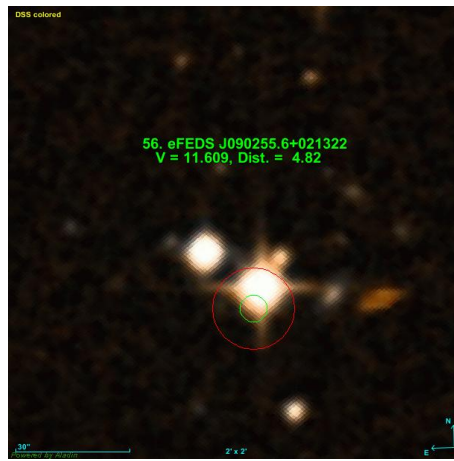
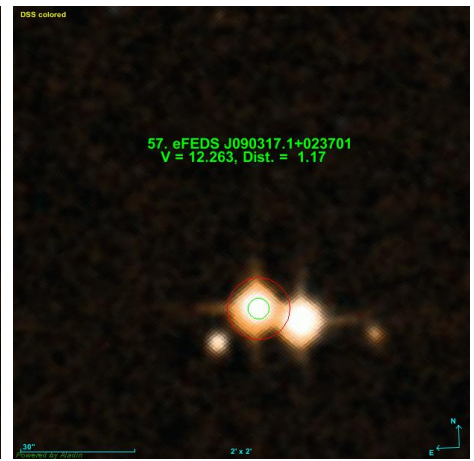

55. eFEDS J090247.7+040023
V = 11.503, Dist. = 1.49

56. eFEDS J090255.6+021322
V = 11.609, Dist. = 4.82

57. eFEDS J090317.1+023701
V = 12.263, Dist. = 1.17

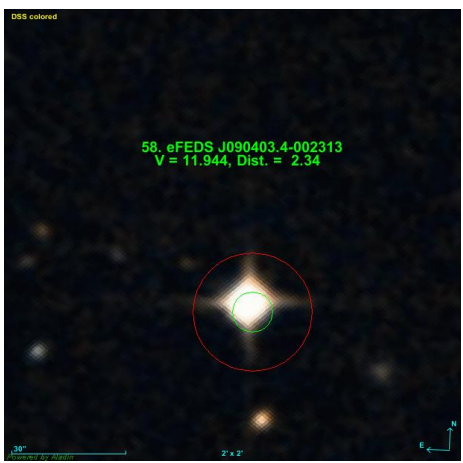
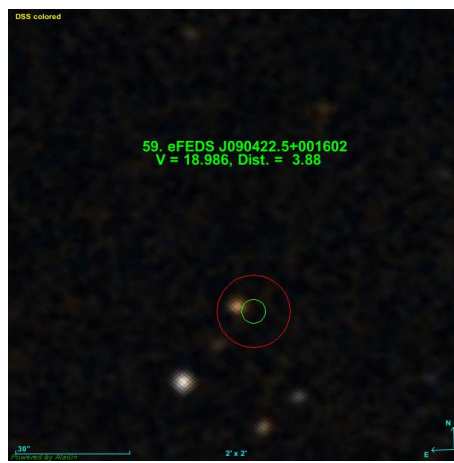
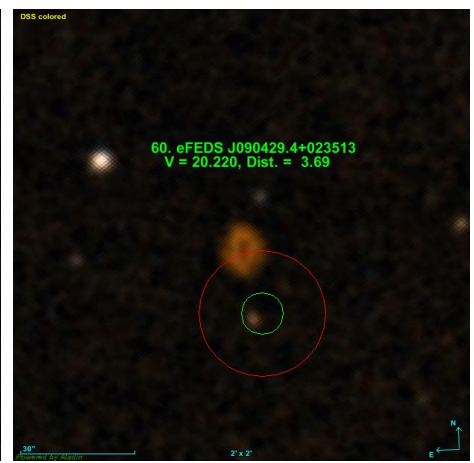

58. eFEDS J090403.4-002313
V = 11.944, Dist. = 2.34

59. eFEDS J090422.5+001602
V = 18.986, Dist. = 3.88

60. eFEDS J090429.4+023513
V = 20.220, Dist. = 3.69

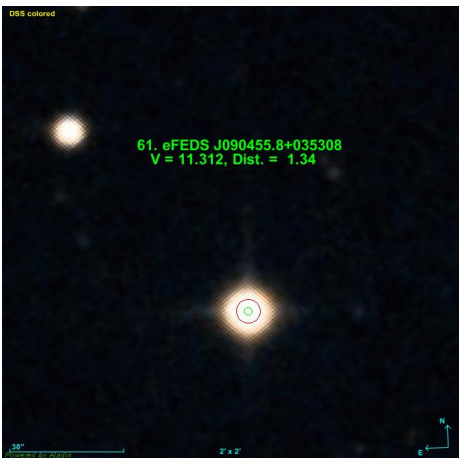

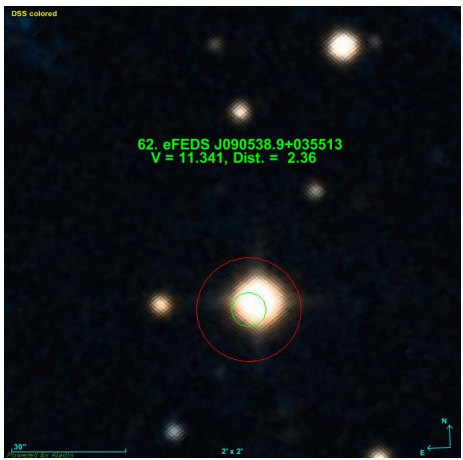

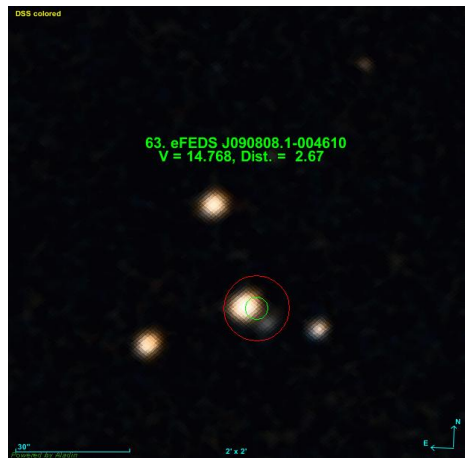

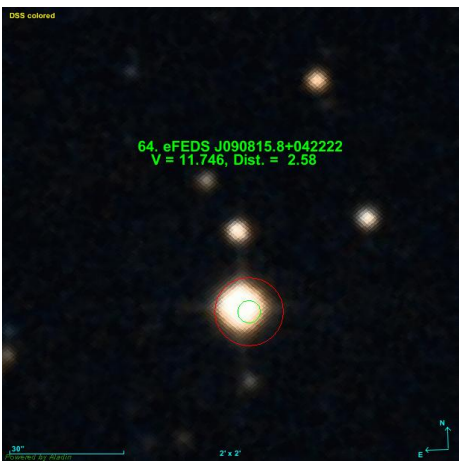

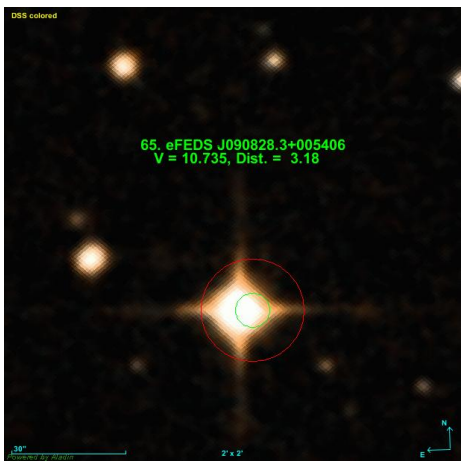

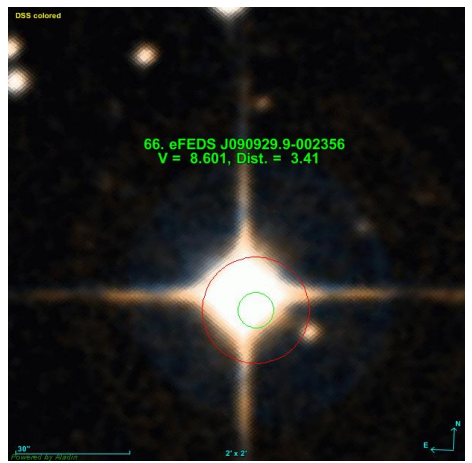

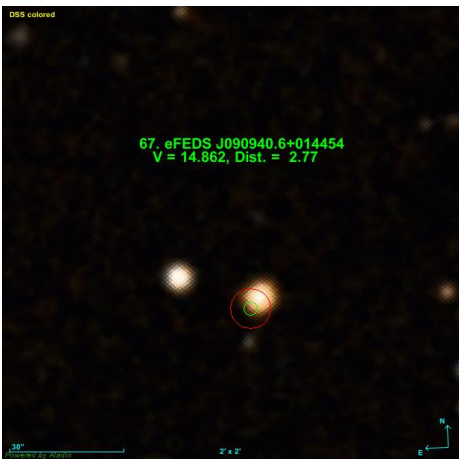

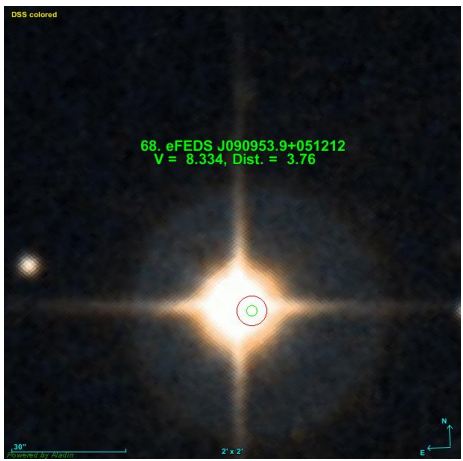

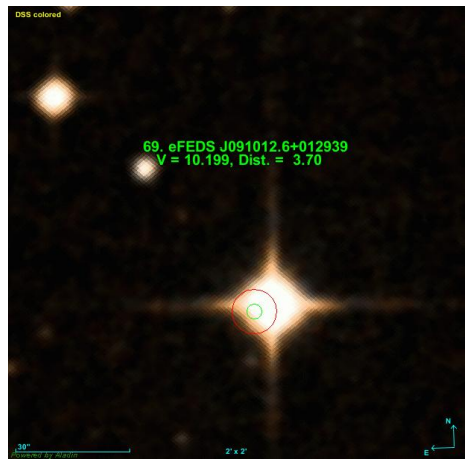

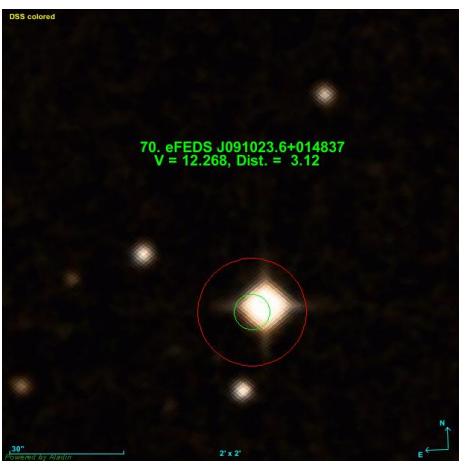

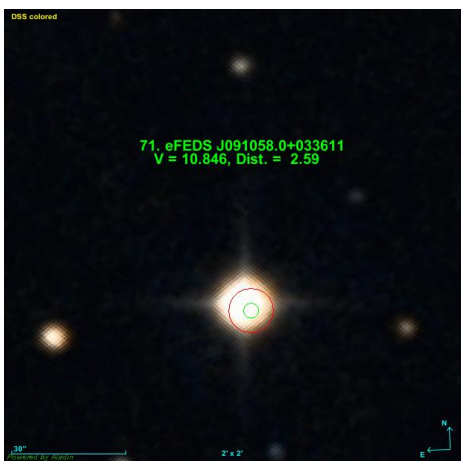

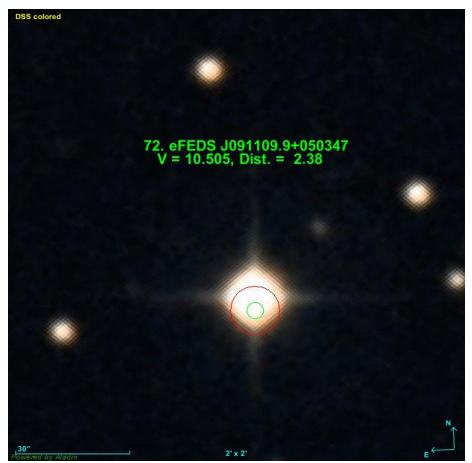

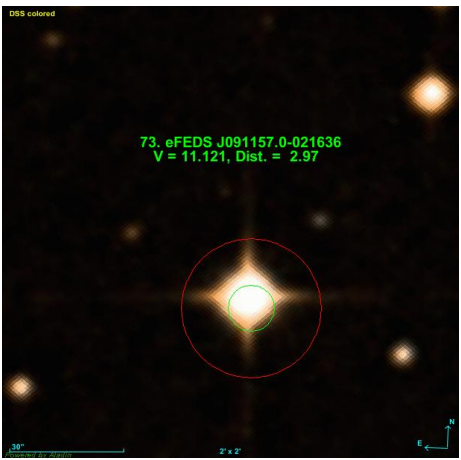

73. eFEDS J091157.0-021636
V = 11.121, Dist. = 2.97

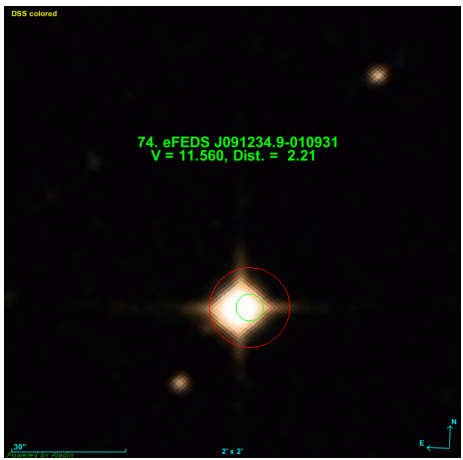

74. eFEDS J091234.9-010931
V = 11.560, Dist. = 2.21

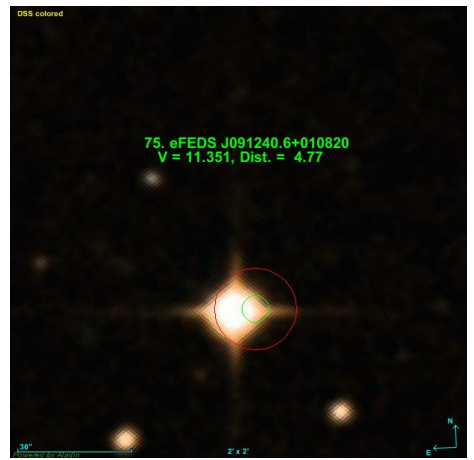

75. eFEDS J091240.6+010820
V = 11.351, Dist. = 4.77

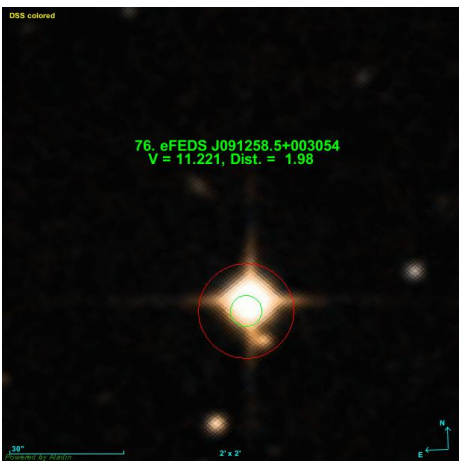

76. eFEDS J091258.5+003054
V = 11.221, Dist. = 1.98

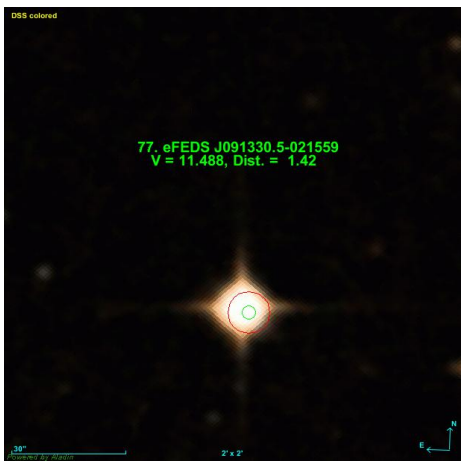

77. eFEDS J091330.5-021559
V = 11.488, Dist. = 1.42

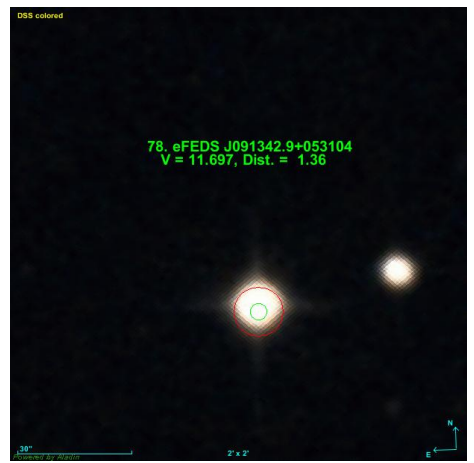

78. eFEDS J091342.9+053104
V = 11.697, Dist. = 1.36

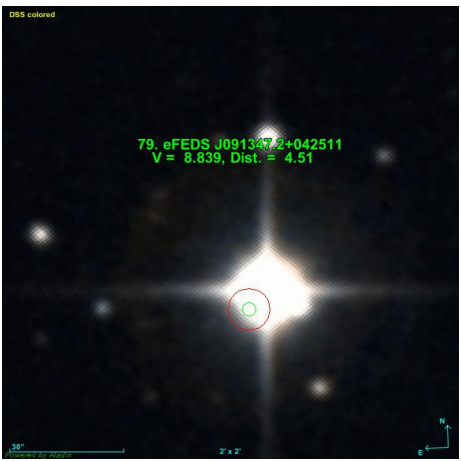

79. eFEDS J091350.2+042511
V = 8.839, Dist. = 4.51

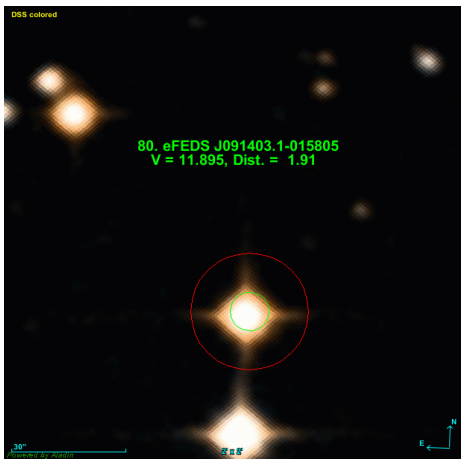

80. eFEDS J091403.1-015805
V = 11.895, Dist. = 1.91

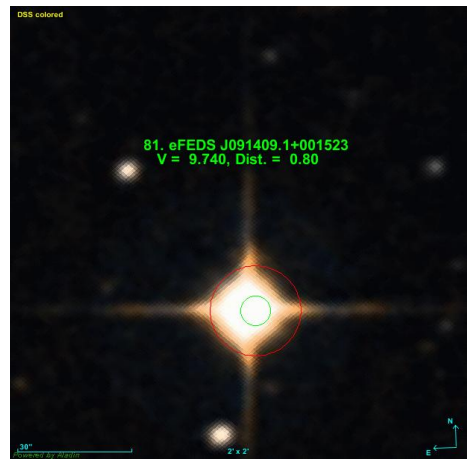

81. eFEDS J091409.1+001523
V = 9.740, Dist. = 0.80

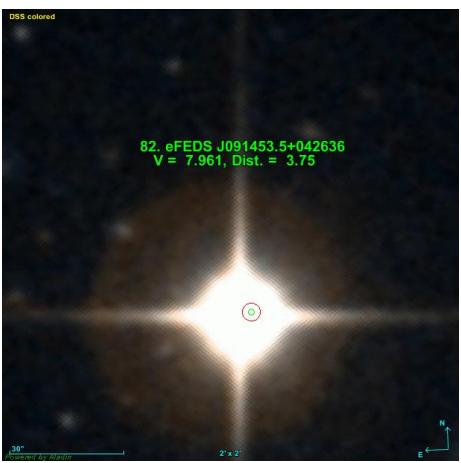

82. eFEDS J091453.5+042636
V = 7.961, Dist. = 3.75

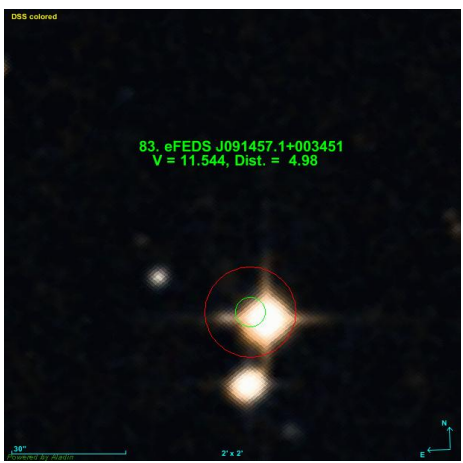

83. eFEDS J091457.1+003451
V = 11.544, Dist. = 4.96

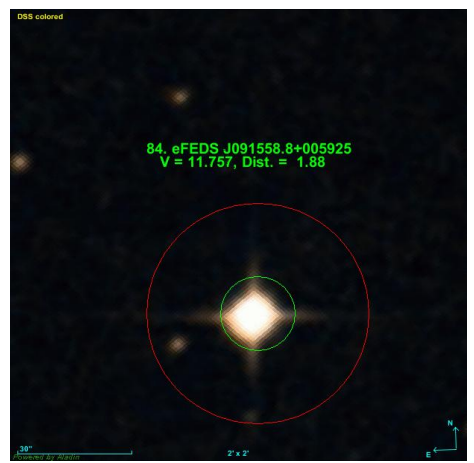

84. eFEDS J091558.8+005925
V = 11.757, Dist. = 1.88

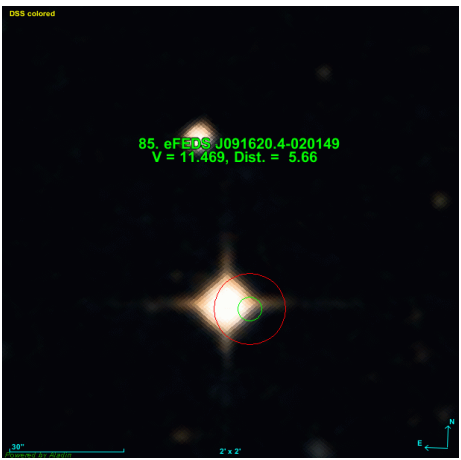

85, eFEDS J091620.4-020149
V = 11.469, Dist. = 5.66

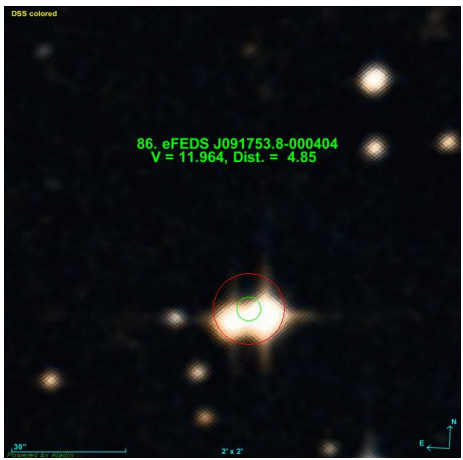

86, eFEDS J091753.8-000404
V = 11.964, Dist. = 4.85

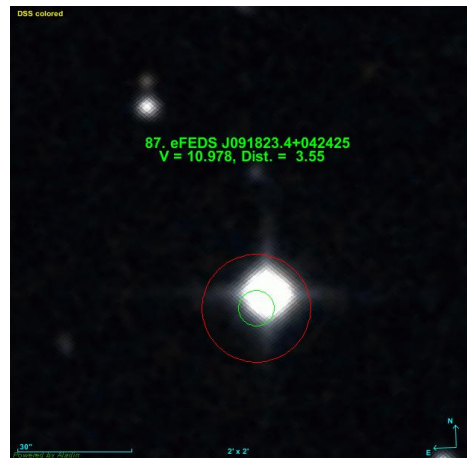

87, eFEDS J091823.4+042425
V = 10.978, Dist. = 3.55

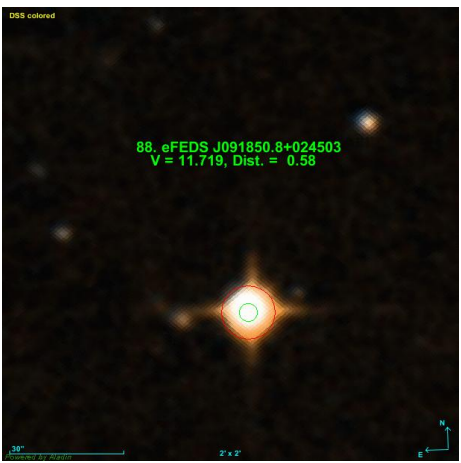

88, eFEDS J091850.8+024503
V = 11.719, Dist. = 0.58

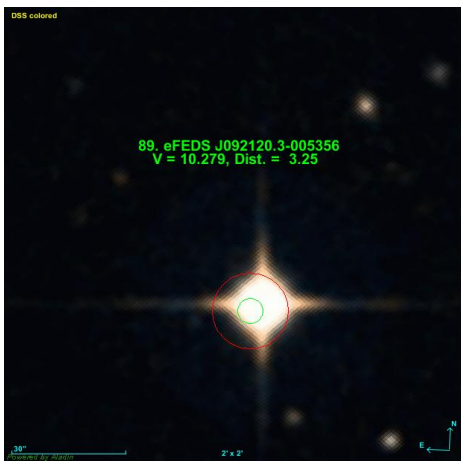

89, eFEDS J092120.3-005356
V = 10.279, Dist. = 3.25

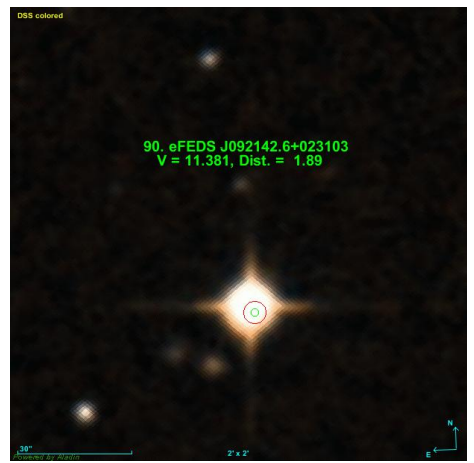

90, eFEDS J092142.6+023103
V = 11.381, Dist. = 1.89

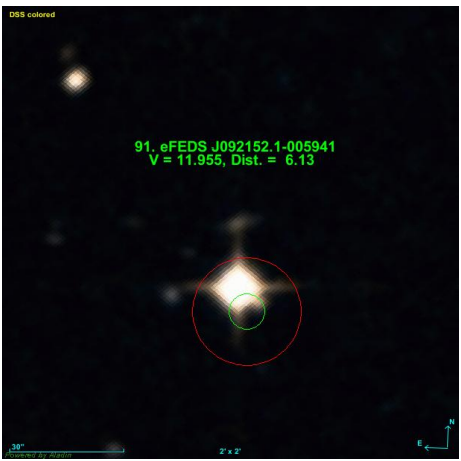

91, eFEDS J092152.1-065941
V = 11.955, Dist. = 6.13

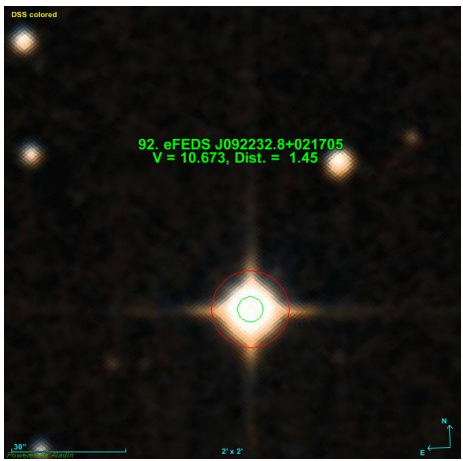

92, eFEDS J092232.8+021705
V = 10.673, Dist. = 1.45

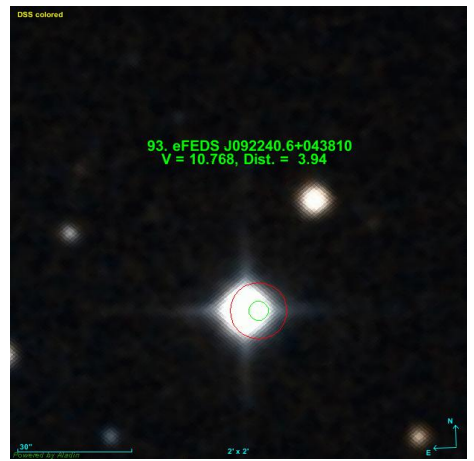

93, eFEDS J092240.6+043810
V = 10.768, Dist. = 3.94

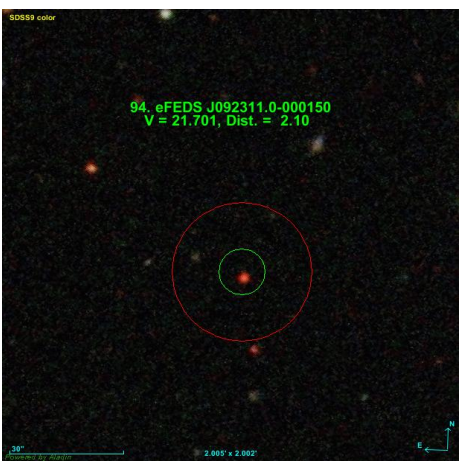

94, eFEDS J092311.0-000150
V = 21.701, Dist. = 2.10

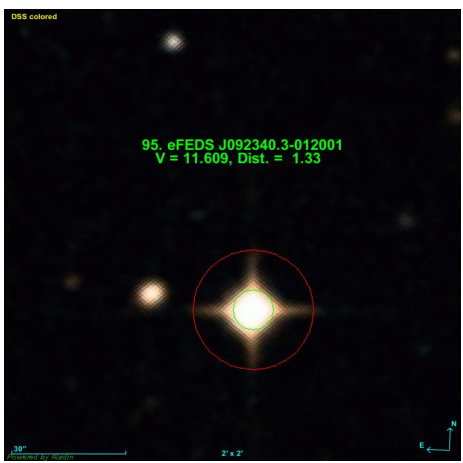

95, eFEDS J092340.3-012001
V = 11.609, Dist. = 1.33

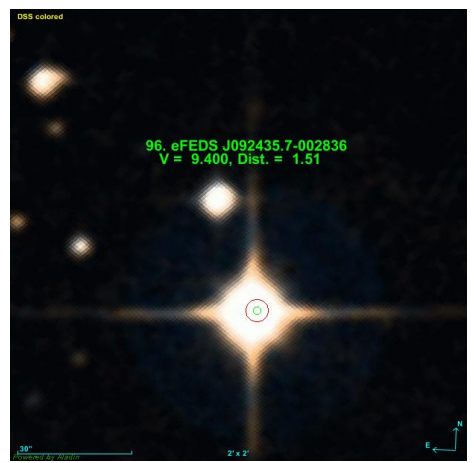

96, eFEDS J092435.7-002836
V = 9.400, Dist. = 1.51

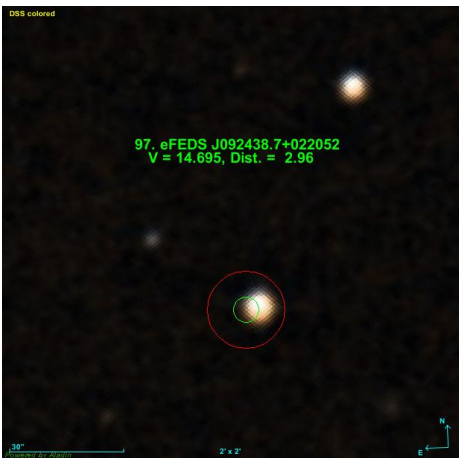
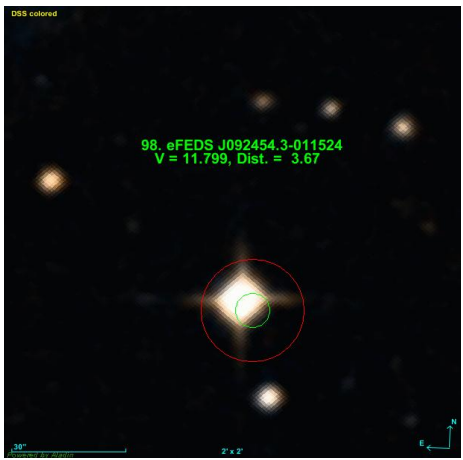
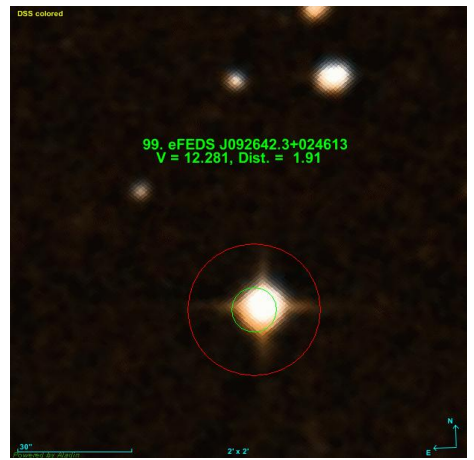

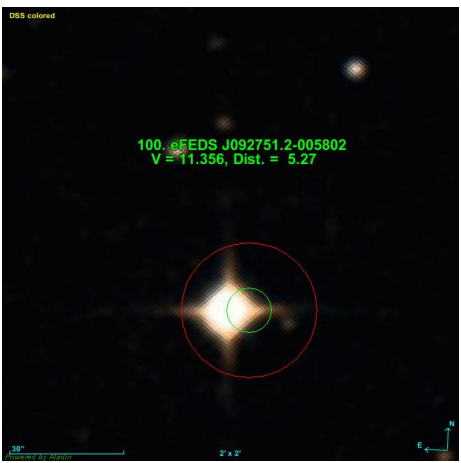
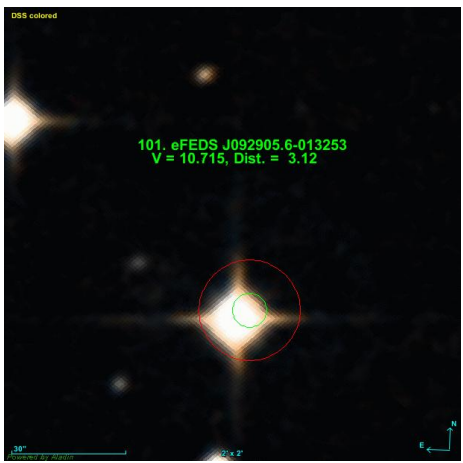
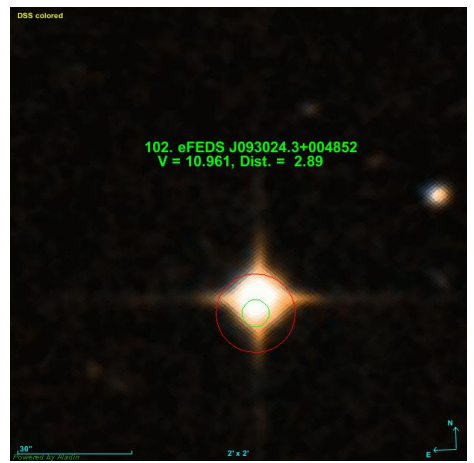

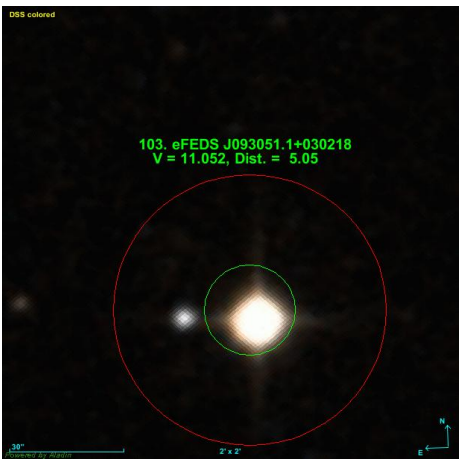
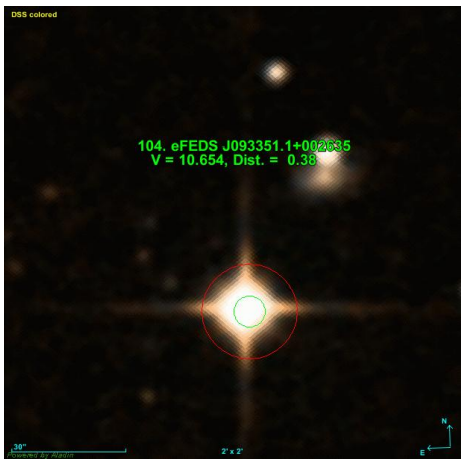
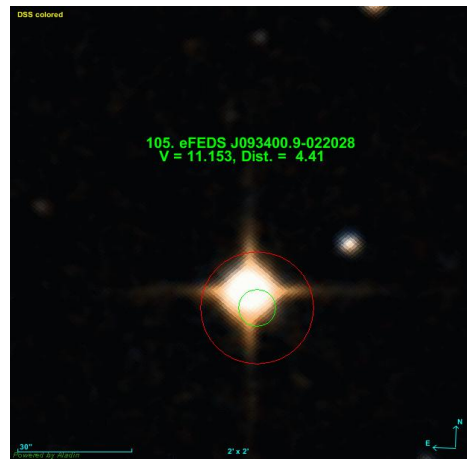

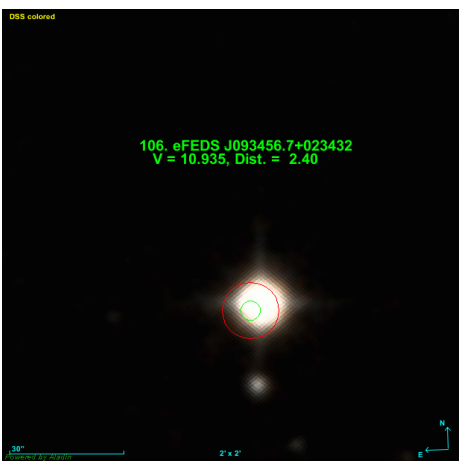
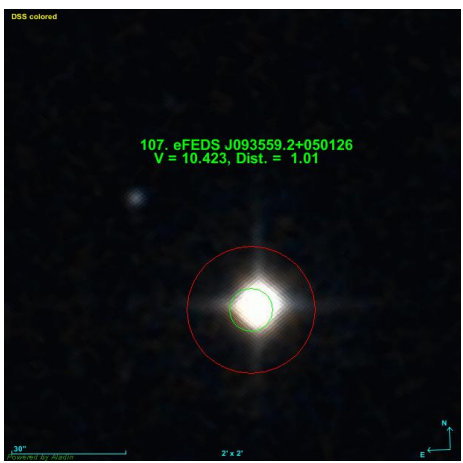
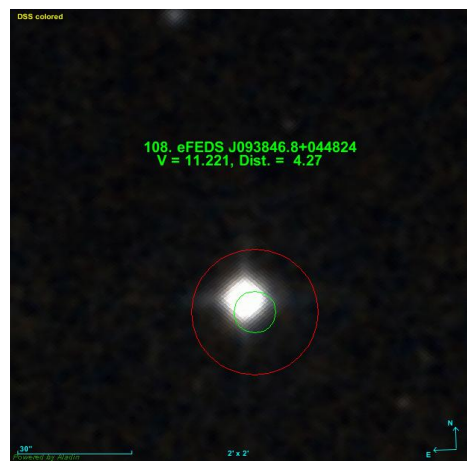

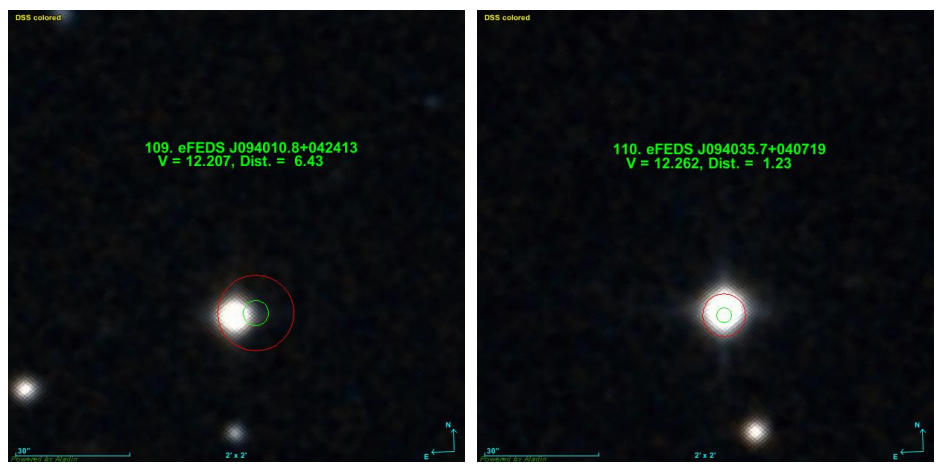

Comments:

№ 5. Two stars in the field: TYC 210-1860-1 and BD+00 2334 (V=9.45, Sp=K0).

№ 6. Faint star near: Gaia DR2 3080242208438978688 (G_mag=18.058).

№ 18. Two faint star near: Gaia DR1: Source ID 3072407707052266752 (top right, G_mag=16.220) and Source ID 3072407707054922112 (below a bright star, G_mag=19.879).

№ 24. Faint star near. Below the bright star on the right (Gaia DR1: Source ID 3075322168782736384, G_mag=16.838).

№ 28. l-m* - low mass star (low-mass*) and №№: 31, 42, 52, 59, 60, 94.

№ 31. Two faint objects in error box: low mass star (SDSS J084817.52+045645.0, G_mag=19.211) and Seyfert 1 Galaxy (SDSS J084818.23+045643.1, G_mag=20.306) - probable identification.

№ 36. Peculiar star near (UCAC4 440-047757, G_mag=13.839).

№ 40. Object below the bright star (Gaia DR1: Source ID 3075652606386315520, G_mag=18.090).

№ 44. Two bright stars in error box: TYC213-1211-1 and Gaia DR2 source ID 576837717289109248 (G_mag=10.593, Teff=5666.60, Rsol=1.18, Lsol=1.297).

№ 52. Close pair of stars: SDSS J090209.49+040742.8 and Gaia DR2 source ID 578617478723879424 (V_mag=17.853, Sp=4.5V, Teff=4305.50).

№ 63. The faint galaxy near (SDSS J090807.97-004613.9, V_mag=19.039). Below the bright star on the right (distance ~ 5 arcsec).

№ 86. Two stars in the field: TYC 4880-182-1 and Gaia DR2 source ID 3842362856369154048 (G_mag=12.866, Teff=4789.52, Rsol=0.70, Lsol=0.233).

№ 90. RoV* - rotation variable (RotV*).

№ 91. Faint star near. Below the bright star on the right (Gaia DR2: Source ID 3839195777550937984, G_mag=18.560).